\documentclass[a4paper,11pt]{article}
\pdfoutput=1 

\usepackage{jheppub} 
\usepackage[T1]{fontenc} 
\usepackage{pdfpages}
\usepackage{amsmath}
\usepackage{braket}
\usepackage{amsfonts}
\usepackage{xcolor}
\usepackage{array}
\usepackage{multirow}
\usepackage{wrapfig}
\usepackage{caption}
\usepackage{subcaption}
\usepackage{tabularx}
\usepackage{placeins}  
\usepackage{mathtools}
\usepackage{amssymb}
\usepackage{lscape}
\usepackage{rotating}
\usepackage{bm}
\usepackage{blindtext}
\usepackage{amsthm}
\usepackage{colortbl}
\usepackage{floatrow}
\usepackage{arydshln}
\usepackage{array}
\usepackage{booktabs}
\usepackage[export]{adjustbox}
\usepackage{multirow,bigdelim}
\usepackage[utf8]{inputenc}

\newcommand\scalemath[2]{\scalebox{#1}{\mbox{\ensuremath{\displaystyle #2}}}}
\newcommand{\overbar}[1]{\mkern 1.5mu\overline{\mkern-1.5mu#1\mkern-1.5mu}\mkern 1.5mu}

\let\Oldsection\section
\renewcommand{\section}{\FloatBarrier\Oldsection}

\newcommand{\threelJ}{\prescript{3\!}{}{\ell}_J}
\newcommand{\threelprimeJ}{\prescript{3\!}{}{\ell'}_{\!\!J}}
\newcommand{\threeSone}{\prescript{3\!}{}{S}_1} 
\newcommand{\threePzero}{\prescript{3\!}{}{P}_0}
\newcommand{\threePone}{\prescript{3\!}{}{P}_1}
\newcommand{\threePtwo}{\prescript{3\!}{}{P}_2}  
\newcommand{\threeDone}{\prescript{3\!}{}{D}_1}
\newcommand{\threeDtwo}{\prescript{3\!}{}{D}_2}
\newcommand{\threeDthree}{\prescript{3\!}{}{D}_3}  
\newcommand{\threeFtwo}{\prescript{3\!}{}{F}_2}
\newcommand{\threeFthree}{\prescript{3\!}{}{F}_3}
\newcommand{\threeFfour}{\prescript{3\!}{}{F}_4}  
\newcommand{\threeGthree}{\prescript{3\!}{}{G}_3}
\newcommand{\threeGfour}{\prescript{3\!}{}{G}_4}

\newcommand{\threeGthreeit}{\prescript{\mathit{3}\!}{}{\mathit{G}}_{\mathit{3}}}
\newcommand{\threeHfourit}{\prescript{\mathit{3}\!}{}{\mathit{H}}_{\mathit{4}}}

\graphicspath{{./figures/}}

\usepackage{natbib}
\title{\boldmath Dynamically-coupled partial-waves in $\rho\pi$ isospin-2 scattering from lattice QCD}

\author[a]{Antoni J. Woss,}
\author[a]{Christopher E. Thomas,}
\author[b,c]{Jozef J. Dudek,}
\author[c]{Robert G. Edwards,}
\author[d]{David J. Wilson}
\author{\\(for the Hadron Spectrum Collaboration)}

\affiliation[a]{DAMTP, University of Cambridge, Centre for Mathematical Sciences, Wilberforce Road, Cambridge CB3 0WA, UK}
\affiliation[b]{Department of Physics, College of William and Mary, 300 Ukrop Way, Williamsburg, VA 23187, USA}
\affiliation[c]{Thomas Jefferson National Accelerator Facility, 12000 Jefferson Avenue, Newport News, VA 23606, USA}
\affiliation[d]{School of Mathematics, Trinity College, Dublin 2, Ireland}

\emailAdd{a.j.woss@damtp.cam.ac.uk}
\emailAdd{c.e.thomas@damtp.cam.ac.uk}
\emailAdd{dudek@jlab.org}
\emailAdd{edwards@jlab.org}
\emailAdd{djwilson@maths.tcd.ie}

\abstract{
We present the first determination of $\rho \pi$ scattering, incorporating dynamically-coupled partial-waves, using lattice QCD, a first-principles numerical approach to QCD. Considering the case of isospin-2 $\rho \pi$, we calculate partial-wave amplitudes with $J \le 3$ and determine the degree of dynamical mixing between the coupled $S$ and $D$-wave channels with $J^P=1^+$.
The analysis makes use of the relationship between scattering amplitudes and the discrete spectrum of states in the finite volume lattice.
Constraints on the scattering amplitudes are provided by over one hundred energy levels computed on two lattice volumes at various overall momenta and in several irreducible representations of the relevant symmetry groups. The spectra follow from variational analyses of matrices of correlations functions computed with large bases of meson-meson operators.
Calculations are performed with degenerate light and strange quarks tuned to the physical strange quark mass so that $m_\pi \sim 700$ MeV, ensuring that the $\rho$ is stable against strong decay.
This work demonstrates the successful application of techniques, opening the door to calculations of scattering processes that incorporate the effects of dynamically-coupled partial-waves, including those involving resonances or bound states.
}

\preprint{\begin{tabular}{r}DAMTP-2018-6\\JLAB-THY-18-2643\end{tabular}}
\arxivnumber{1802.05580}

\begin{document} 

\maketitle
\flushbottom


\section{Introduction \label{Sec:Introduction}}

Hadron spectroscopy is predominantly the investigation of resonances which decay strongly into hadrons, such as the pion, which are stable under the strong interaction. Many resonances which decay into multi-meson final states do so through an intermediate state featuring resonances of non-zero intrinsic spin. For example, the axial-vector $a_1(1260)$ meson dominantly decays into a $\pi\pi\pi$ final state through $\rho(770) \pi$, where the vector $\rho(770)$ decays into $\pi \pi$. Once an intermediate hadron has non-zero intrinsic spin, it becomes possible for more than one \mbox{partial-wave} to be present for a given $J^P$ through the coupling of the orbital angular momentum $\ell$ to the intrinsic spin $S$. For example, in the case of the $J^P=1^+$ $a_1$ decaying to $\rho\pi$, where the $\rho$ has $S=1$, both $S$ and $D$-waves can contribute, and indeed it is possible to measure the relative decay amplitudes~\cite{Link:2007fi}.

QCD is the theory of the strong interaction which confines quarks and gluons inside hadrons and leads to residual interactions between hadrons. The confining property makes calculations of QCD at low energies very difficult and a convenient approach, which allows the theory to be attacked numerically, is to consider QCD on a finite Euclidean lattice of space-time points. This formulation, known as lattice QCD, has been applied to compute energy spectra and other quantities of interest in hadron spectroscopy from correlation functions. The spectra so extracted are discrete owing to the finite spatial size of the lattice. Well below the threshold for strong decay, the discrete energies correspond to the energies of stable hadrons. More generally, it has been shown that infinite-volume hadron scattering amplitudes can be related to the finite-volume spectra through a quantisation condition derived originally by L\"{u}scher~\cite{Luscher:1985dn,Luscher:1986pf} and subsequently extended by many others~\cite{Bedaque:2004kc,Bernard:2010fp,Briceno:2014oea,Briceno:2012yi,Briceno:2013hya,Christ:2005gi,Feng:2004ua,Fu:2011xz,Guo:2012hv,Hansen:2012tf,Kim:2005gf,Lage:2009zv,Leskovec:2012gb,Li:2012bi,He:2005ey,Rummukainen:1995vs}. 

While significant progress has been made studying meson-meson scattering using lattice QCD~\cite{Briceno:2017max}, calculations have not accounted for the effects of dynamically-coupled \mbox{partial-waves} when processes feature scattering hadrons with non-zero intrinsic spin\footnote{Some recent work which has considered \mbox{vector-pseudoscalar} scattering in the light sector and makes brief comment on the possibility of contributions from \mbox{dynamically-coupled} \mbox{partial-waves}, but does not incorporate this in the analysis, can be found in Ref.~\cite{Lang:2014tia}.}. It is to this problem that we turn here.

Nucleon-nucleon scattering in the spin-triplet channel has the same \mbox{partial-wave} decomposition as $\rho\pi$ scattering, and a closely related quantisation condition in finite-volume\footnote{There is a slightly smaller symmetry in $\rho\pi$ owing to the unequal masses of the $\rho$ and the $\pi$.}. A non-relativistic quantisation condition for $NN$ was presented in \cite{Briceno:2013bda}, and an attempt to determine the $\threeSone, \threeDone$ mixing appeared in \cite{Orginos:2015aya}.

In this paper we report on the first calculation of the energy dependence of \mbox{partial-wave} scattering amplitudes for $\rho\pi$ in isospin-2, including the coupled $S$ and $D$-wave system with $J^P=1^+$. In this exploratory study, we work with heavier-than-physical light quarks, so the $\rho$ becomes a stable hadron lying some way below the $\pi\pi$ threshold. Specifically, we work at the $\text{SU}(3)$ flavour symmetric point with three degenerate flavours of quark ($u,d,s$) tuned to have mass approximately equal to the physical strange quark mass, leading to a pion mass $\sim 700\,\text{MeV}$. In this way we are justified in considering elastic $\rho\pi$ scattering provided we stay below the $\pi\pi\pi$ threshold\footnote{No complete formalism for relating finite-volume spectra to three-body scattering amplitudes yet exists, but see~\cite{Briceno:2012rv,Hansen:2014eka,Hansen:2015zga,Polejaeva:2012ut,Briceno:2017tce,Hammer:2017kms,Mai:2017bge,Doring:2018xxx} for progress.}.

The exotic isospin considered here leads us to expect that the $\rho\pi$ scattering amplitudes will be non-resonant and, based upon experience taken from $\pi\pi$ scattering, they are likely to be relatively weak. A study of $\rho\pi$ scattering within a non-relativistic quark model~\cite{Barnes:2000hu} found weak, mainly repulsive scattering, with the $\threeSone$ phase-shift being largest, but not exceeding $-35^\circ$, and a rather small mixing between the $\threeSone$ and $\threeDone$ \mbox{partial-waves}.   

The weakness of the $\rho\pi$ interactions in isospin-2 will lead to small shifts in energy in the finite-volume spectrum with respect to the energies expected were $\rho$ and $\pi$ to have no residual hadron-hadron interactions. The small energy shifts must be accurately and reliably calculated.  This can be achieved by employing a large basis of interpolating operators, $\mathcal{O}_i$, having the quantum numbers of isospin-2 $\rho\pi$, to calculate a matrix of correlation functions,
\begin{equation}\label{Cij}
C_{ij}(t)=\braket{0|\mathcal{O}_i(t)\mathcal{O}_j^{\dagger}(0)|0} \, ,
\end{equation}
and a variational analysis~\cite{Michael:1985ne,Luscher:1990ck} can then be applied to reliably extract the energy spectrum.

For the case of vector-pseudoscalar scattering, the total intrinsic spin $S=1$ can couple with the orbital angular momentum $\ell$ to give three distinct total angular momenta $J$ for $\ell \geq 1$. In the absence of interactions, this gives rise to many degenerate energy levels -- these may only be split slightly in the interacting case. A large operator basis containing appropriate operator structures is essential in order to disentangle these \mbox{near-degenerate} states.

We utilise the relevant symmetries of the finite volume when calculating correlation functions which allows us to identify which \mbox{partial-waves} are contributing to each energy level. In a limited number of cases, an energy level is dominantly affected by a single \mbox{partial-wave}, and here a value of the phase-shift for that \mbox{partial-wave}, at that energy, can be determined via a one-to-one mapping. More generally, an energy level is affected by multiple \mbox{partial-waves} and a more sophisticated analysis technique is required -- the energy dependence of \mbox{partial-wave} amplitudes is parameterised and multiple energy levels are considered simultaneously. This approach is similar to that used in coupled-channel cases~\cite{Dudek:2016cru,Moir:2016srx,Wilson:2014cna,Wilson:2015dqa}. Significant constraints on scattering amplitudes come from spectra computed for systems with overall non-zero momentum with respect to the lattice, and indeed we find that the sign of the off-diagonal coupling between $S$-wave and $D$-wave can only be obtained from such `\mbox{in-flight}' cases. We begin by examining the features of vector-pseudoscalar scattering in an infinite volume.

\section{Vector-pseudoscalar scattering \label{Sec:Vector_Pseudoscalar_Scattering}}
In this section, we discuss the features of a scattering process that involves one or more hadrons with non-zero intrinsic spin. We explore the consequences for hadron-hadron scattering in an infinite volume and distinguish these from features that are purely a consequence of the finite volume. The results are illustrated by a discussion of vector-pseudoscalar scattering.

\subsection{Infinite Volume}\label{infvolsec}

In an \mbox{infinite-volume} continuum, total angular momentum $J$ is a good quantum number and can be constructed by taking a tensor product of the orbital angular momentum $\ell$ with the total intrinsic spin $S$ (itself constructed via a tensor product of the spins of the two scattering hadrons), i.e.\ ${\ell \otimes S  = |\ell - S| \oplus ... \oplus \ell + S}$. Parity, $P$, is another good quantum number and is given by $P=\eta_1\eta_2(-1)^\ell$, where $\eta_1$ and $\eta_2$ are the intrinsic parities of the hadrons. It follows that, in some cases, hadron-hadron states with a particular $J^P$ can be formed from \emph{multiple} $\ell S$ combinations\footnote{The choice of the $\ell S$ basis as opposed to, say, a helicity basis is one made for later convenience: it has the advantage that the threshold behaviour of $\ell S$ basis states is given in terms of the value of $\ell$.}.

For the case of vector-pseudoscalar scattering, $S=1$, and thus, for $\ell \geq 1$, $J$ can take one of a triplet of values $J=\{\ell-1,\ell,\ell+1\}$. The intrinsic parities of vector and pseudoscalar mesons are each negative and it follows that $J^P=1^+,\,2^-,\,3^+...$ can each be formed from \emph{two} distinct $\ell S$ combinations. In spectroscopic notation, $^{2S+1}\ell_J$, these are $\{\threeSone,\,\threeDone\}$, $\{\threePtwo,\,\threeFtwo\}$, $\{\threeDthree,\,\threeGthree\}\,, ...$ respectively. For these $J^P$ values, even though the scattering process may only have a single hadron-hadron channel kinematically open, there are two \mbox{partial-wave} channels which can couple dynamically. For example, considering $J^P = 1^+$, the $\mathbf{t}$-matrix\footnote{related to the unitary $\mathbf{S}$-matrix by $\mathbf{S}=\mathbf{1}+2i\rho\,\mathbf{t}$} can be written as,
\begin{align}\label{tinf}
\mathbf{t} &= 
\begin{bmatrix}
t(\threeSone|\threeSone) & t(\threeSone|\threeDone) \\ 
t(\threeSone|\threeDone) & t(\threeDone|\threeDone)
\end{bmatrix} \nonumber \\
&= \frac{1}{2 i \rho} 
\begin{bmatrix}
\cos(2\bar{\epsilon})  \exp\big[2 i\, \delta_{\threeSone}\big] - 1 & i \sin(2\bar{\epsilon})  \exp\big[i(\delta_{\threeSone} + \delta_{\threeDone})\big] \\
i \sin(2\bar{\epsilon})  \exp\big[i(\delta_{\threeSone} + \delta_{\threeDone})\big] & \cos(2\bar{\epsilon})  \exp\big[2 i\, \delta_{\threeDone}\big] - 1
\end{bmatrix}, 
\end{align}
where $\rho(E_\mathsf{cm}) = 2 k_\mathsf{cm}/E_\mathsf{cm}$ is the phase-space factor and the second line presents the common Stapp parameterisation \cite{Stapp:1956mz} in terms of two \emph{phase-shifts}, $\delta_{\threeSone}(E_\mathsf{cm})$, $\delta_{\threeDone}(E_\mathsf{cm})$, and a \emph{mixing angle}, $\bar{\epsilon}(E_\mathsf{cm})$, describing the coupling between the two channels\footnote{The sign of the off-diagonal entries, and hence the sign of $\bar{\epsilon}$, is physically relevant and impacts the spin and angular dependence of the scattering amplitudes. This is in contrast to the case where different hadronic channels are coupled -- there the sign cannot be measured and it is usual to parameterise in terms of an \emph{inelasticity} parameter which discards this sign information.}. The symmetric nature of the $\mathbf{t}$-matrix follows from the time-reversal symmetry of QCD. This parameterisation automatically respects coupled-channel unitarity, expressed in this context as $\mathrm{Im}\, [t^{-1}({\threelJ}|{\threelprimeJ})] = -\rho\,\delta_{\ell\ell'}$ for energies above threshold, where the phase-space is the same for both the $\threeSone$ and $\threeDone$ channels\footnote{When there are additional coupled channels featuring different scattering hadrons, $\rho(E_\mathsf{cm})$ is diagonal in the channel space but no longer proportional to the identity as $k_\mathsf{cm}$ depends on the scattering hadron masses.}. Within the $\ell S$ basis, the threshold behaviour of the $\mathbf{t}$-matrix elements is simple: $t\big({\threelJ} | {\threelprimeJ}\big) \propto \big(k_\mathsf{cm}\big)^{\ell+\ell'}$.

\subsection{Finite Volume}\label{secFV}

Lattice calculations like the ones we report on in this article are performed in a finite periodic cubic volume, and this causes there to be `mixing' between \mbox{partial-waves} that cannot mix dynamically in an infinite volume. This is a consequence of the broken rotational symmetry caused by working in an $L\times L\times L$ volume. For systems overall at rest, the symmetry is reduced to that of the double cover of the octahedral group $\text{O}_h^{\text{D}}$. The infinite-volume irreducible representations (\emph{irreps}), labelled $\big(J,m\big)$ where $m$ is the projection of $J$ along the $z$-axis, get \emph{subduced} into the finite number of irreps of $\text{O}_h^{\text{D}}$, labelled $\big(\Lambda,\mu \big)$ with $\Lambda$ the irrep and $\mu$ the row within that irrep\footnote{The rows, $\mu$, are analogous to the projections, $m$, in the rotationally symmetric case. In this work we will consider only the integer-spin irreps, relevant for \mbox{meson-meson} scattering, arising from the single cover of $\text{O}_h$.}. As such, multiple \mbox{partial-waves} of distinct $J$ can populate the same irrep -- in fact an infinite number can. We summarise the subduction of low-lying \mbox{partial-waves} of the vector-pseudoscalar system in Table~\ref{tab000}. The subduction is controlled only by values of $J^P$, but recall from the discussion above that in some cases multiple $\threelJ$ constructions can give the same $J^P$ -- the table distinguishes these two possible types of `mixing'.

For systems with non-zero overall momentum $\vec{P}$, the periodic boundary conditions on the spatial volume restrict $\vec{P}$ to a discrete set of values given by $\vec{P}=\frac{2\pi}{L}\vec{n}$ where $\vec{n}=(n_x,n_y,n_z)$ with $n_i \in \mathbb{Z}$. We use a shorthand notation when labelling momenta in which the $2\pi/L$ factor is omitted, e.g.\ $\vec{P}=[n_x,n_y,n_z]$ or $[n_xn_yn_z]$. These `\mbox{in-flight}' systems have a symmetry which is further reduced and can be described by the \emph{little group}, $\text{LG}(\vec{P})$, the subgroup of $\text{O}_h^{\text{D}}$ that leaves $\vec{P}$ invariant, and this reduced symmetry leads to a subduction pattern that is more dense in $J$ values. Furthermore, parity is no longer a good quantum number. A more complete discussion of the little groups can be found in Ref.~\cite{Moore:2005dw}. For $|\vec{n}|^2\leq 4$, the \mbox{partial-wave} subductions for a vector-pseudoscalar system are presented in Tables~\ref{tab001} -~\ref{tab111} in Appendix \ref{App:Subduction_In_Flight}. 

\begin{table}[tb] 
{\renewcommand{\arraystretch}{1.2}
	\centering
\begin{tabular}{ c |l l l l l} 
	\centering
$\Lambda^+$ & \multicolumn{1}{c}{$A_1^+$} &  \multicolumn{1}{c}{$A_2^+$} &  \multicolumn{1}{c}{$T_1^+$} &  \multicolumn{1}{c}{$E^+$} &  \multicolumn{1}{c}{$T_2^+$} \\ 
\hline 
&&&&&\\
\multirow{4}{15mm}{$J^+(\threelJ)$}&&&   $1^+ \left( \begin{matrix} \prescript{3\!}{}{S}_1 \\  \prescript{3\!}{}{D}_1 \end{matrix}   \right)$  & &\\
&&&& $2^+ \,\,\left(\prescript{3\!}{}{D}_2 \right)$ & $2^+ \,\,\left(\prescript{3\!}{}{D}_2 \right)$  \\
&& $3^+ \left(\begin{matrix} \prescript{3\!}{}{D}_3 \\  \prescript{3}{}{G}_3 \end{matrix} \right)$ 
& $3^+ \left(\begin{matrix} \prescript{3\!}{}{D}_3 \\  \prescript{3}{}{G}_3 \end{matrix}\right)$ 
&& $3^+ \left(\begin{matrix} \prescript{3\!}{}{D}_3 \\  \prescript{3}{}{G}_3 \end{matrix} \right)$ \\
& $4^+ \,\,\left(\prescript{3}{}{G}_4 \right)$ && $4^+ \,\,\left(\prescript{3}{}{G}_4 \right)$ & $4^+ \,\,\left(\prescript{3}{}{G}_4 \right)$ & $4^+\,\, \left(\prescript{3}{}{G}_4 \right)$ \\
\hline
\multicolumn{6}{c}{\vspace{0cm}}\\
$\Lambda^-$ & \multicolumn{1}{c}{$A_1^-$} &  \multicolumn{1}{c}{$A_2^-$} &  \multicolumn{1}{c}{$T_1^-$} &  \multicolumn{1}{c}{$E^-$} &  \multicolumn{1}{c}{$T_2^-$} \\ 
\hline 
\multirow{7}{15mm}{$J^-(\threelJ)$}& $0^-\,\, \left(\prescript{3\!}{}{P}_0 \right)$ &&&&\\
& & & $1^- \,\,\left(\prescript{3\!}{}{P}_1 \right)$ &&\\
&&&& $2^- \left( \begin{matrix} \prescript{3\!}{}{P}_2 \\  \prescript{3\!}{}{F}_2 \end{matrix}   \right)$ &
$2^- \left( \begin{matrix} \prescript{3\!}{}{P}_2 \\  \prescript{3\!}{}{F}_2 \end{matrix}   \right)$ \\
& & $3^- \,\,\left(\prescript{3\!}{}{F}_3 \right)$ & $3^-\,\, \left(\prescript{3\!}{}{F}_3 \right)$ && $3^- \,\,\left(\prescript{3\!}{}{F}_3 \right)$ \\
& $4^- \left( \begin{matrix} \prescript{3\!}{}{F}_4 \\  \prescript{3\!}{}{H}_4 \end{matrix}   \right)$ &&
$4^- \left( \begin{matrix} \prescript{3\!}{}{F}_4 \\  \prescript{3\!}{}{H}_4 \end{matrix}   \right)$ &
$4^- \left( \begin{matrix} \prescript{3\!}{}{F}_4 \\  \prescript{3\!}{}{H}_4 \end{matrix}   \right)$ &
$4^- \left( \begin{matrix} \prescript{3\!}{}{F}_4 \\  \prescript{3\!}{}{H}_4 \end{matrix}   \right)$ \\
\hline
\end{tabular} 
\caption{Subduction of \mbox{partial-waves}, $\threelJ$, for $J\leq 4$ into the irreps, $\Lambda^{P}$, of the octahedral group, $\text{O}_h$, relevant for systems overall at rest. The notation $J^P({\threelJ})$ denotes the \mbox{partial-wave} content for a given $J^P$, with multiple $\threelJ$ entries indicating \mbox{partial-waves} which mix dynamically.  This table is derived from Table 2 of \cite{Johnson:1982yq}.}
\label{tab000}
}
\end{table} 

In order to determine infinite-volume scattering amplitudes, we calculate finite-volume energy levels and utilise a quantisation condition, first derived by L\"{u}scher~\cite{Luscher:1985dn,Luscher:1986pf,Luscher:1990ux,Luscher:1991cf}, which relates the two quantities. If, in a certain energy region, only one \mbox{partial-wave} has a non-negligible value, the relation takes the commonly-used form
\begin{equation}\label{elas_luescher}
\text{cot} \, \delta(E_\mathsf{cm}) = -\text{cot}\,\phi(E_\mathsf{cm},L)\,, 
\end{equation}
where $\phi(E_\mathsf{cm},L)$ is a known function that encodes the kinematical and symmetry-breaking effects of the finite volume. In this case, each \mbox{finite-volume} energy level can be used to determine the value of the \mbox{partial-wave} phase-shift at that particular energy. In the case of vector-pseudoscalar scattering, an example might be the rest-frame $E^+$ irrep at energies near threshold. Here the $\threeDtwo$ wave is expected to be much larger than the $\threeGfour$ wave, or any wave of still higher $\ell$, owing to the effect of the centrifugal barrier which ensures that $t(\threeDtwo|\!\threeDtwo) \sim (k_\mathsf{cm})^4 \gg t(\threeGfour|\!\threeGfour) \sim (k_\mathsf{cm})^8$. If multiple energy levels can be obtained, from calculations on one or more volumes at rest and \mbox{in-flight}, repeated use of Eq.~(\ref{elas_luescher}) will yield the energy-dependence of the phase-shift\footnote{A demonstration of this can be seen in $\pi\pi$ isospin-1 scattering in $P$-wave -- see Figure 10 in Ref.~\cite{Dudek:2012xn}.}. 

Where multiple \mbox{partial-waves} are present, but still only a single hadron-hadron channel is kinematically accessible, the L\"{u}scher quantisation condition for a given irrep can be summarised by an equation,
\begin{equation}\label{full_luescher}
\det \Big[ \mathbf{1} + i \rho(E_\mathsf{cm})\,  \mathbf{t}(E_\mathsf{cm}) \cdot \big( \mathbf{1} + i \overline{ \bm{\mathcal{M}} }(E_\mathsf{cm},L) \big) \Big] = 0 \,,
\end{equation}
where the determinant is over all \mbox{partial-waves} subduced into that irrep. For a known \mbox{$\mathbf{t}$-matrix}, the zeros of the determinant give the discrete spectrum $\big\{E^{(k)}_{\mathsf{cm}}(L) \big\}$ in an $L\times L \times L$ box. The \mbox{$\mathbf{t}$-matrix} respects the symmetries of the infinite volume and is therefore diagonal in $J$, while $\overline{\bm{\mathcal{M}}}$ is a matrix, dense in the space of \mbox{partial-waves}, of known functions of $E_\mathsf{cm}$ and box size $L$, encoding the effects of the finite volume. In the case of only a single \mbox{partial-wave} being significant, $\mathbf{t}$ and $\overline{\bm{\mathcal{M}}}$ are $1\times 1$ matrices, and Eq.~(\ref{full_luescher}) reduces to Eq.~(\ref{elas_luescher}) -- see Appendix~\ref{App:Details_Of_Luescher} for more details. 

Eq.~(\ref{full_luescher}) encodes both the dynamical mixing of \mbox{partial-waves} (present even in an infinite volume), through $\mathbf{t}$, and the `mixing' of \mbox{partial-waves} due to the finite volume, through $\overline{\bm{\mathcal{M}}}$. For example, in the rest-frame $T_1^+$ irrep, considering the \mbox{partial-wave} content with $\ell\leq 2$, we have dynamical mixing between the $\threeSone$ and $\threeDone$-waves with $J^P=1^+$. The $J^P=3^+$ wave $\threeDthree$ `mixes' with $1^+$ only because of the reduced symmetry of the finite volume. The $\mathbf{t}$-matrix is
\begin{equation}\label{tinffin}
\mathbf{t} = 
\begin{bmatrix}
 t(\threeSone |\! \threeSone) & t(\threeSone |\! \threeDone) & 0 \\
 t(\threeSone |\! \threeDone) & t(\threeDone |\! \threeDone) & 0 \\
 0 & 0 & t(\threeDthree |\! \threeDthree)
\end{bmatrix} ,
\end{equation}
where the off-diagonal contributions dynamically couple $\threeSone$ and $\threeDone$. The non-vanishing elements of $\overline{\bm{\mathcal{M}}}$ in this $3 \times 3$ space ensure that all three waves contribute in the determination of the \mbox{finite-volume} spectrum.

In the case of multiple \mbox{partial-waves}, coupled either dynamically or due to the finite volume, each energy level provides a constraint on the $\mathbf{t}$-matrix at that energy, through Eq.~(\ref{full_luescher}), but use of one such equation is not sufficient to determine the multiple unknowns in $\mathbf{t}$. A number of such constraints, each coming from a different finite-volume energy level, are required to determine $\mathbf{t}(E_\mathsf{cm})$. Considering systems with overall non-zero momentum is one way to obtain many energy levels -- the moving frame changes the spatial boundary conditions, which in turn modifies the quantisation condition giving a different set of functions in $\overline{\bm{\mathcal{M}}}$. This is discussed in detail in Ref. \cite{Rummukainen:1995vs,Christ:2005gi,Kim:2005gf} and has been successfully applied in determinations of coupled-channel $\mathbf{t}$-matrices in Refs.~\cite{Dudek:2016cru,Moir:2016srx,Wilson:2014cna,Wilson:2015dqa,Briceno:2017qmb,Dudek:2014qha}. We will present the details of the approach, relevant to the current case of vector-pseudoscalar scattering, in Section~\ref{Sec:Results}.

\section{Spectrum determination \label{Sec:Spectrum_Determination}}
To make a robust determination of the finite-volume energy spectrum in each irrep, we compute an $N\times N$ matrix of two-point correlation functions using $N$ independent interpolating operators with appropriate quantum numbers, $C_{ij}(t) = \big\langle 0 \big| \mathcal{O}^{}_i(t+t_{\text{src}}) \, \mathcal{O}^\dag_j(t_{\text{src}}) \big| 0 \big\rangle$. We extract the spectra using the \emph{variational method}~\cite{Michael:1985ne}, applying an implementation detailed in Refs.~\cite{Dudek:2010wm,Dudek:2007wv} as used in numerous works~\cite{Dudek:2012gj,Dudek:2012xn,Wilson:2014cna,Wilson:2015dqa,Dudek:2016cru,Briceno:2017qmb,Briceno:2016mjc,Moir:2016srx,Cheung:2017tnt,Dudek:2011tt,Liu:2012ze,Woss:2016tys}. 

In brief, the approach is to solve the generalised eigenvalue problem,
\begin{equation}\label{gevp}
C^{}_{ij}(t) \, v^{(n)}_j=\lambda_n(t,t_0) \, C^{}_{ij}(t_0) \, v^{(n)}_j \,,
\end{equation}
where the $n^{\textrm{th}}$ eigenvalue $\lambda_n(t,t_0)$, also known as the $n^{\textrm{th}}$ \emph{principal correlator}, contains information about the energy of the $n^{\textrm{th}}$ state $E_n$, and the eigenvector $v^{(n)}$ provides the optimal linear combination of the basis of $N$ operators to interpolate the $n^{\textrm{th}}$ state. We choose an appropriate $t_0$ as explained in Ref.~\cite{Dudek:2007wv} and check robustness of the determined spectrum by considering a range of $t_0$'s. Energies are obtained by fitting principal correlators to the form $\lambda_n(t,t_0)=(1-A_n)\, e^{-E_n(t-t_0)}+A_n \, e^{-E'_n(t-t_0)}$, where $A_n, E_n'$ parameterise the small residual excited state pollution and are not used further.

The optimal operator to interpolate the $n^{\textrm{th}}$ eigenstate is given by $\Omega_n^\dagger = \sum_i v^{(n)}_i \mathcal{O}_i^\dagger$. These optimised operators relax to the $n^{\textrm{th}}$ state at earlier times than any one single operator in the basis; an example of the improvement for the ground state pion at various momenta can be seen in Figure 2 of Ref.~\cite{Dudek:2012gj}. We discuss the use of optimised \mbox{single-meson} operators in the construction of \mbox{meson-meson} interpolating operators in Section~\ref{tw_me_op}. 

In order to investigate \mbox{meson-meson} scattering, we need to construct an appropriate set of operator structures which overlap strongly onto the eigenstates of QCD in a finite volume with the quantum numbers of the \mbox{meson-meson} scattering problem. Operators which resemble \mbox{meson-meson} states, constructed as products of operators which resemble single mesons of definite momentum, prove to be very effective -- see e.g.\ Figure 6 of Ref.~\cite{Wilson:2015dqa}. We describe how to construct these \mbox{meson-meson} operators in the sections to follow, with a particular focus relevant to this calculation on operators that respect $\text{SU}(3)_{\text{F}}$ flavour symmetry and which resemble vector-pseudoscalar states.

\subsection{Single-meson operators in $\text{SU}(3)$ flavour representations}\label{si_me_op}
Following Refs.~\cite{Dudek:2010wm,Thomas:2011rh}, we construct \emph{\mbox{single-meson} operators} from fermion bilinears.  These have a spin and spatial structure built from Dirac $\gamma$-matrices and gauge-covariant derivatives, are projected onto overall momentum $\vec{p}$, and have a flavour structure that transforms in a particular $\text{SU}(3)_{\text{F}}$ multiplet. Schematically the construction is,
\begin{equation}\label{qbilinears}
{\mathcal{O}^\dagger}^{Jm}_{\bm{F}\nu}(\vec{p}, t) = \sum_{\vec{x}}e^{i\vec{p}\cdot \vec{x}}\,\sum_{\nu_1, \, \nu_2} \mathcal{C}\!\left( \begin{array}{ccc}
{\bm{\bar{3}}} & {\bm{3}} & {\bm{F}} \\
\nu_1 & \nu_2 & \nu \end{array} \right){\bar{q}}_{\nu_1}(\vec{x},t) \, \mathbf{\Gamma}_t   \,{q}_{\nu_2}(\vec{x}, t)\,.
\end{equation}
Here $\mathbf{\Gamma}_t$ denotes a product of $\gamma$-matrices and up to $3$ gauge-covariant derivatives acting in position space, colour and Dirac spin-space on time-slice $t$. The constructions are engineered to have definite continuum $J^P$ and $m$ where, for $\vec{p}=\vec{0}$, $m$ is the projection of $J$ along the $z$-axis and, for $\vec{p}\neq \vec{0}$, $m$ is replaced by the \emph{helicity}, $\lambda$ -- see Ref.\ \cite{Thomas:2011rh}. The quark fields, $q_{\nu}(\vec{x},t)$, corresponding to the up, down and strange quarks $(u,d,s)$, are in the $\bm{3}$ multiplet of $\text{SU}(3)_{\text{F}}$ -- the elements can be labelled by $\nu=(I,Y,I_z)$, where $I$ is the isospin, $Y$ is the hypercharge and $I_z$ is the $z$-component of isospin. The $\mathcal{C}(...)$ are $\text{SU}(3)_{\text{F}}$ Clebsch-Gordan coefficients following conventions given in De Swart~\cite{deSwart:1963pdg}, and the sum over $\text{SU}(3)_{\text{F}}$ components projects the quark-bilinear onto a definite $\text{SU}(3)_{\text{F}}$ flavour multiplet $\bm{F}$, which can be either $\mathbf{1}$ or $\mathbf{8}$. When $Y=0$, these operators have definite $G$-parity\footnote{$G$-parity is a generalisation of charge-conjugation, $C$, where the $G$-parity operator, $\hat{G}=\hat{C}e^{i\pi \hat{I}_y}$ is a rotation of $\pi$ around the $I_y$ component of isospin followed by the charge-conjugation operation.}.

These operators of definite $J^P$ and $m$ are subduced into the appropriate lattice irreps of $\mathrm{O}_h$ or $\text{LG}(\vec{p})$ as discussed in Ref.\ \cite{Thomas:2011rh}. The subduction does not impact the flavour representation and the result is an operator, 
${\mathcal{O}^\dagger}^{\Lambda \mu}_{\bm{F} \nu}(\vec{p},t) = \sum_m \mathcal{S}^{J m}_{\Lambda \mu}\,\, {\mathcal{O}^\dagger}^{Jm}_{\bm{F}\nu}(\vec{p},t)$, 
in a particular irrep. Tabulated values of the subduction coefficients, $\mathcal{S}^{J m}_{\Lambda \mu}$, for $\vec{p}=\vec{0}$ can be found in the Appendix of Ref.~\cite{Dudek:2010wm}, and for $\vec{p}\neq\vec{0}$ in Table II of~\cite{Thomas:2011rh}.

As an example, consider a pseudoscalar $\text{SU}(3)_{\text{F}}$ singlet, $\bm{F}=\bm{1}$, $\nu=(0,0,0)$, $\bm{\Gamma}_t=\gamma_5$ and $\vec{p}=\vec{0}$. Subducing Eq.\ (\ref{qbilinears}) gives the operator,
\[
\mathcal{O}^{\dagger A_1^-\,1}_{\,\,\bm{1}\,(0,0,0)}=\tfrac{1}{\sqrt{3}}(\bar{u}\gamma_5 u + \bar{d}\gamma_5 d + \bar{s}\gamma_5 s) \,.
\]
\subsection{Meson-meson operators in $\text{SU}(3)$ flavour representations}\label{tw_me_op}
Operators which resemble a pair of mesons can be constructed from a product of two \mbox{single-meson} operators. Our approach here follows that presented in Refs.~\cite{Dudek:2012gj,Dudek:2012xn}, and in this section we will concentrate on constructing operators in definite $\text{\text{SU}}(3)_{\text{F}}$ multiplets. Writing out the flavour structure explicitly, the \mbox{meson-meson} operator takes the form,       
\begin{equation}\label{MM}
{\mathcal{O}^\dagger}^{\Lambda \mu}_{\bm{F} \nu}  \Big( \substack{\bm{F_1}\Lambda_1 \vec{p}_1 \\ \bm{F_2}\Lambda_2 \vec{p}_2 } \Big| \vec{P}  \Big)
 = \sum_{\nu_1,\,\nu_2}
 \mathcal{C}\!\left( \begin{array}{ccc}
{\bm{F_1}} & {\bm{F_2}} & {\bm{F}} \\
\nu_1 & \nu_2 & \nu \end{array} \right)
\sum_{\mu_1,\,\mu_2}
\mathbb{C}\!\left( \begin{array}{ccc}
\Lambda_1& \Lambda_2 & \Lambda \\
\mu_1 & \mu_2 & \mu \end{array} \right)
\sum_{\substack{ \vec{p}_i \in \{\vec{p}_i\}^*\\ \vec{p}_1+\vec{p}_2 = \vec{P}}} 
{\Omega^\dagger}^{\Lambda_1 \mu_1}_{\bm{F_1} \nu_1}(\vec{p}_1)\;
{\Omega^\dagger}^{\Lambda_2 \mu_2}_{\bm{F_2} \nu_2}(\vec{p}_2),
\end{equation}
where the optimised operator ${\Omega^\dagger}^{\Lambda_i \mu_i}_{\bm{F_i}\nu_i}(\vec{p}_i)$ interpolates a meson of momentum $\vec{p}_i$ in the $\bm{F_i}$ flavour multiplet with component $\nu_i$. `Lattice' Clebsch-Gordan coefficients, $\mathbb{C}(...)$, are required to couple irreps $\Lambda_1 \otimes \Lambda_2 \rightarrow \Lambda$, and the momentum sum runs over all momenta related to $\vec{p}_i$ by an allowed lattice rotation, $\vec{p}_i \in \{\vec{p}_i\}^*$, such that $\vec{p}_1 + \vec{p}_2 = \vec{P}$ -- see Ref.~\cite{Dudek:2012gj} for details.

Since \mbox{single-meson} operators are restricted to the $\text{SU}(3)_{\text{F}}$ octet, $\bm{8}$, and singlet, $\bm{1}$, \mbox{meson-meson} operators are restricted to the $\bm{27},\bm{10},\bm{\overbar{10}},\bm{8}$ and $\bm{1}$ multiplets. In this work, we will perform calculations with exact $\text{SU}(3)_{\text{F}}$ symmetry and focus on $I=2$ $\rho\pi$ scattering which lies in the $\bm{27}$ multiplet. We are at liberty to choose any component of the $\bm{27}$ multiplet when we calculate the energy spectra, as they are all equivalent, and we choose ${\nu=(I=2,Y=0,I_z=2)}$. The $\text{SU}(3)_{\text{F}}$ Clebsch-Gordan coefficients in Eq.~(\ref{MM}) ensure that the relevant \mbox{meson-meson} operators come from products of \mbox{single-meson} operators with flavour structure $\bm{F}=\bm{8}$ and $\nu=(1,0,1)$. $G$-parity ensures that there are no pseudoscalar-pseudoscalar or vector-vector channels which can mix with $I=2$ $\rho\pi$.

The basis of \mbox{meson-meson} operators used to form the matrix $C_{ij}(t)$ can be constructed using different magnitudes of momentum\footnote{Strictly speaking, this should be momentum `type' or star, $\{\vec{p}_i\}^*$ as indicated in Eq.~(\ref{MM}), rather than magnitude, but the distinction is not relevant for the momenta considered in this paper.}, $\big|\vec{p}_1 \big|, \big|\vec{p}_2\big|$, where directions of the momenta are summed over in Eq.~(\ref{MM}) subject to $\vec{p}_1 + \vec{p}_2 = \vec{P}$. There is a close association between the finite-volume energy-levels when mesons have no meson-meson interactions, ${E_\mathsf{n.i.} = \sqrt{m_\pi^2 + |\vec{p}_1|^2 } + \sqrt{m_\rho^2 + |\vec{p}_2|^2 }}$, which we refer to as `\mbox{non-interacting}' energies, and these operators. Earlier studies have found that \mbox{meson-meson} operators which closely resemble the \mbox{non-interacting} states in the energy range of interest are efficient at interpolating \mbox{finite-volume} correlation functions~\cite{Dudek:2012gj,Dudek:2012xn}. This suggests that, if we are interested in only a certain energy range, operators which correspond to a \mbox{non-interacting} energy which lies far above this energy region do not need to be included in the basis.

When a \mbox{single-meson} operator for a vector meson has non-zero momentum, the reduced symmetry of the lattice means that the different helicity components subduce into $N_{\lambda}$ different irreps of $\text{LG}(\vec{p}_1)$. Each of these vector operators can be combined, via Eq.~(\ref{MM}), with a pseudoscalar operator transforming in some irrep of $\text{LG}(\vec{p}_2)$, to form a set of linearly-independent vector-pseudoscalar operators at some overall momentum $\vec{P}$ in some irrep $\Lambda$. Furthermore, each vector operator when combined with a pseudoscalar operator may appear numerous times within a single irrep, e.g.~$[001]E_2 \otimes [011]A_2 \rightarrow 2\times [001]E_2$, and form multiple linearly-independent vector-pseudoscalar operators -- we refer to this number as the \emph{multiplicity} (which could be zero). Together, this means that there can be many linearly-independent vector-pseudoscalar operators, transforming within some irrep $\Lambda$, which correspond to the same \mbox{non-interacting} energy and we denote the total number of such operators as $N_{\text{lin}}$. It is important to emphasise that $N_{\text{lin}}$ is the sum of the multiplicities for each of the $N_{\lambda}$ vector operators combined with the appropriate pseudoscalar operator.

For example, consider vector-pseudoscalar operators overall at rest, $\vec{P} = \vec{0}$, in the $T_1^+$ irrep, which we write as $[000]T_1^+$. The operator corresponding to lowest \mbox{non-interacting} energy features a vector meson at rest (in the $T_1^-$ irrep) coupled to a pseudoscalar at rest (in the $A_1^-$ irrep). In this case, $N_{\lambda}=1$, and there is only one operator corresponding to the one way of coupling ${[000]T_1^- \otimes [000]A_1^-\to [000]T_1^+}$ ($N_{\text{lin}}=1$). Of course, there are still three equivalent rows of the $T_1$ irrep.

On the other hand, for a vector meson with momentum $\vec{p}=[001]$, the helicity $0$ and $\pm1$ components subduce into the $[001]A_1$ and $[001]E_2$ irreps respectively ($N_\lambda=2$). Combining the vector with a pseudoscalar so that the vector-pseudoscalar operator is overall at rest, there are \emph{two} linearly independent operators transforming in $[000]T_1^+$ from ${[001]A_1\otimes [001]A_2 \rightarrow [000]T_1^+}$ and ${[001]E_2\otimes [001]A_2 \to [000]T_1^+}$ ($N_{\text{lin}}=2$). 

If the vector meson has momentum $\vec{p}=[011]$, the three helicities subduce into three different irreps, $[011]A_1$, $[011]B_1$ and $[011]B_2$ ($N_\lambda=3$). When combined appropriately with the pseudoscalar, this gives \emph{three} linearly-independent vector-pseudoscalar operators transforming in $[000]T_1^+$ from ${[011]A_1 \otimes [011]A_2 \rightarrow [000]T_1^+}$, ${[011]B_1 \otimes [011]A_2 \rightarrow [000]T_1^+}$ and ${[011]B_2 \otimes [011]A_2 \rightarrow [000]T_1^+}$ ($N_{\text{lin}}=3$). 

While we have illustrated how multiple \mbox{meson-meson} operators with the same associated \mbox{non-interacting} energies can arise by considering a vector-pseudoscalar operator overall at rest, this situation also occurs when there is an overall non-zero momentum.  For example, with $\vec{P} = [001]$, ${[001]A_1 \otimes [011]A_2 \to [001]E_2}$ and ${[001]E_2 \otimes [011]A_2 \to 2\times [001]E_2}$ giving $N_\lambda=2$ and $N_{\text{lin}}=3$ (as $[001]E_2 \otimes [011]A_2$ into $[001]E_2$ has a multiplicity of two). In all cases, the \mbox{non-interacting} \mbox{meson-meson} spectrum will feature degeneracies: for each \mbox{non-interacting} energy, the degeneracy is equal to $N_{\text{lin}}$ of the corresponding \mbox{meson-meson} operator. As one might anticipate, failing to include all the occurrences of \mbox{meson-meson} operators in a given energy region can lead to an incomplete spectrum\footnote{See for example \cite{Prelovsek:2014swa} where e.g. only one of the two possible linearly-independent $\psi[001] \pi[001]$ operators, and only one of the three possible linearly-independent $\psi[011] \pi[011]$ operators are included in a calculation of the $[000]T_1^+$ spectrum of hidden charm $I=1$. The resulting spectrum does not have a distribution of energy levels commensurate with the expected degeneracy pattern. The presence of multiple linearly-independent vector-pseudoscalar operators was later recognised in~\cite{Prelovsek:2016iyo}.}. This is demonstrated clearly in Figure 8 of~\cite{Cheung:2017tnt}.

\section{Lattice setup \label{Sec:Lattice_Calculations}}
Calculations of correlation functions were performed on anisotropic lattices of volumes $(L/a_s)^3 \times (T/a_t) = 20^3 \times 128$ and $24^3 \times 128$, with spatial lattice spacing $a_s \sim 0.12\text{ fm}$ and temporal lattice spacing ${a_t = a_s/\xi \sim ( 4.7 \, \mathrm{GeV} )^{-1}}$ where $\xi\sim 3.5$ is the anisotropy. $L$ and $T$ are the spatial and temporal extents of the lattice respectively. We use gauge fields generated from a tree-level Symanzik improved gauge action and a Clover fermion action with $N_f=3$ degenerate flavours of dynamical quarks~\cite{Edwards:2008ja,Lin:2008pr}, tuned to have masses approximately equal to the physical strange quark mass, giving exact $\text{SU}(3)_F$ symmetry. The flavour octet of pseudoscalars has a mass $\sim 700\,\text{MeV}$, while the vector octet has a mass $\sim 1020\,\text{MeV}$. With these heavy masses, exponentially suppressed finite-volume and thermal effects are negligible ($m_\pi L \gtrsim 10$, $m_\pi T \gtrsim 18$). 

For calculating correlation functions we employ \emph{distillation}~\cite{Peardon:2009gh}. This enables us to efficiently compute correlators involving a large basis of operators with various structures by projecting the quark fields to a low-energy subspace (distillation space) of small rank, $N_{\text{vecs}}$. We increase the statistical precision by averaging correlation functions over a number, $N_{\text{tsrcs}}$, of independent time-sources, $t_{\text{src}}$. To reduce statistical correlations between the energy levels for different moving frames, we averaged over a different set of time-sources for each non-zero momentum. The rank of the distillation space, number of gauge configurations, and number of time-sources used for the computations of $\rho$, $\pi$ and $\rho\pi$ correlation functions, on each lattice, are shown in Table \ref{dist_vecs}. 

When quoting results in physical units, we set the scale using the $\Omega$-baryon mass. From the value obtained on a lattice identical to those discussed above but of smaller spatial volume ${(L/a_s)^3 \times (T/a_t) = 16^3 \times 128}$, ${a_tm_\Omega^\textrm{latt.}=0.3593(7)}$~\cite{Edwards:2012fx}, and the experimental mass, ${m_\Omega^{\textrm{exp.}}=1672.45(29)\,\text{MeV}}$ \cite{Patrignani:2016xqp}, we obtain the inverse temporal spacing via ${a_t^{-1}=m_\Omega^{\textrm{exp.}}/a_tm_\Omega^{\textrm{latt.}}}$, giving ${a_t^{-1} = 4655\,\text{MeV}}$.
\begin{table}
	\centering
	\begin{subtable}{.5\textwidth}
		\centering
		\setlength\tabcolsep{1.5pt} 
	\begin{tabular}{c | c c c}
	$(L/a_s)^3 \times (T/a_t)$ & $N_{\text{vecs}}$ & $N_{\text{cfgs}}$ & $N_{\text{tsrcs}}$ \\
	\hline
	$20^3 \times 128$ & 128 & 197 & 8\\
	$24^3 \times 128$ & 160 & 499 & 1\\
\end{tabular}
\label{a}
\caption{}
	\end{subtable}%
	\begin{subtable}{.5\textwidth}
	\centering
	\setlength\tabcolsep{1.5pt} 
	\begin{tabular}{c | c c c}
		$(L/a_s)^3 \times (T/a_t)$ & $N_{\text{vecs}}$ & $N_{\text{cfgs}}$ & $N_{\text{tsrcs}}$ \\
		\hline
		$20^3 \times 128$ & 128 & 502 & 1-3\\
		$24^3 \times 128$ & 160 & 607 & 1-3\\
	\end{tabular}
\label{b}
	\caption{}
\end{subtable}%
\caption{Number of distillation vectors ($N_{\text{vecs}}$), gauge configurations ($N_{\text{cfgs}}$) and time-sources ($N_{\text{tsrcs}}$) used to compute correlation functions on the two lattice volumes, as described in the text, for (left) $\rho$ and $\pi$ correlation functions ($\bm{F}=\bm{8}$) and (right) $\rho\pi$ correlation functions ($\bm{F}=\bm{27}$).}
\label{dist_vecs}
\end{table}

\section{Dispersion relations of the $\rho$ and $\pi$ mesons \label{Sec:Disp_Rel}}
In preparation for studying $\rho\pi$ scattering, we first compute the momentum dependence of the relevant stable mesons' energies, check that they satisfy the relativistic dispersion relations and determine the anisotropy, $\xi\equiv a_s/a_t$. The relativistic dispersion relation for a stable hadron is, up to discretisation corrections,
\begin{equation}\label{disp_rel}
(a_t E_{\vec{n}})^2 = (a_t m)^2 + \frac{1}{\xi^2}\bigg(\frac{2\pi}{L/a_s}\bigg)^2 |\vec{n}|^2  \, ,
\end{equation}
where $m$ is the mass of the hadron and $E_{\vec{n}}$ is its energy with momentum $\vec{p}=\frac{2\pi}{L}\vec{n}$. Differences between the values of $\xi$ measured from different hadrons are due to discretisation, finite volume and/or thermal effects but we expect all but the first of those to be negligible as discussed in Section~\ref{Sec:Lattice_Calculations}. The energies of the ground-state flavour octet vector and pseudoscalar mesons, hereafter referred to as $\rho$ and $\pi$, with momentum $|\vec{n}\,|^2 \leq 4$ were calculated from a variational analysis of matrices of correlation functions involving bases of \mbox{single-meson} operators. The analyses also gave the optimised operators for interpolating the $\rho$ and $\pi$ with the various momenta -- these are used in the construction of vector-pseudoscalar operators as discussed in Section~\ref{Sec:Spectrum_Determination}.

\begin{figure}
	\vspace*{-1cm}
  \includegraphics[scale=0.955]{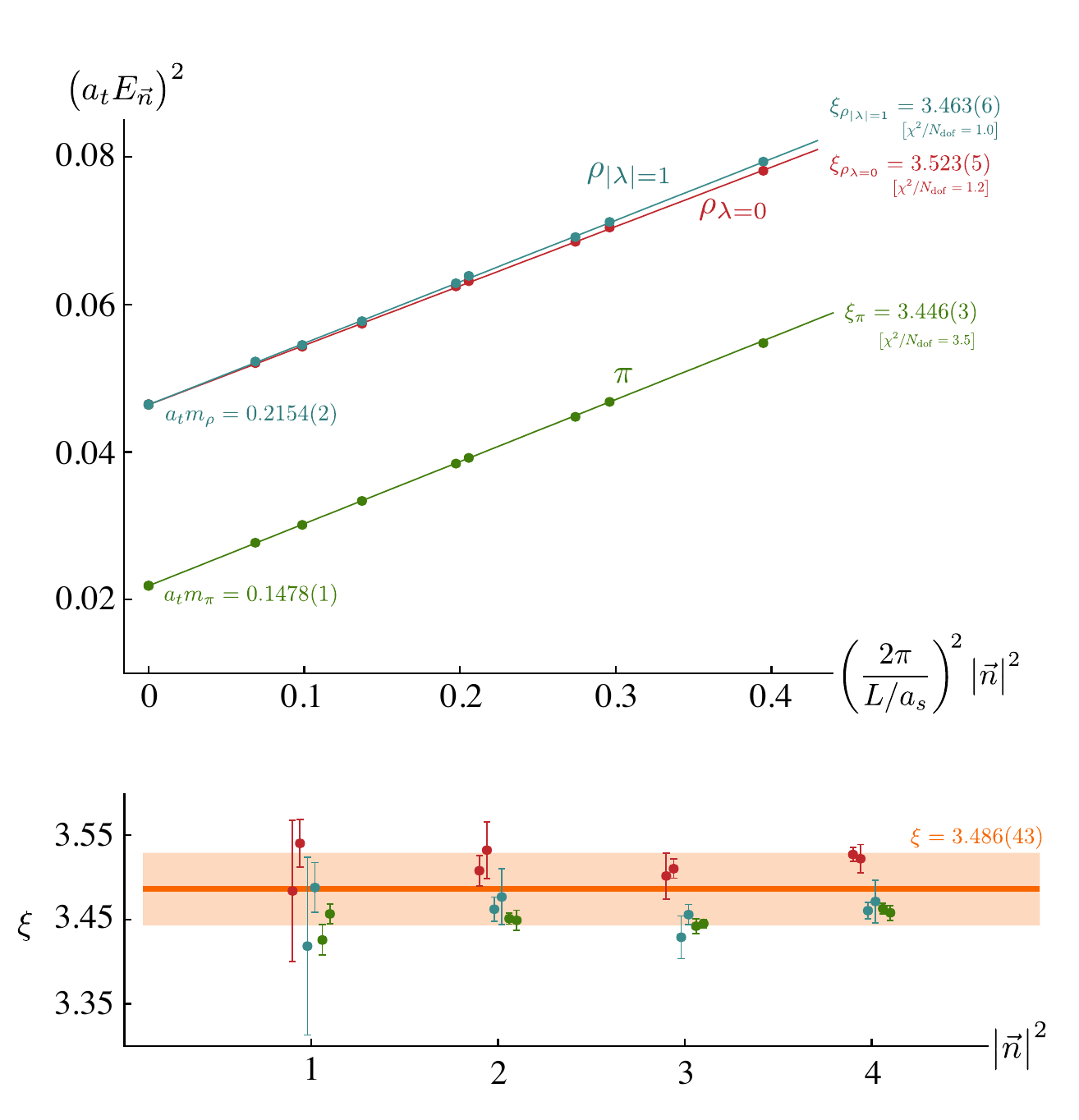}
  \caption{Upper panel: dispersion relations for the $\pi$ and $|\lambda| = 0, 1$ helicity components of the $\rho$. The statistical errors on the energies are smaller than the points. Lines and numerical values show results of fits to determine $\xi$ using Eq.~(\ref{disp_rel}). Lower panel: points show the effective momentum-dependent $\xi$ obtained via $ \left[ \big(\frac{2\pi}{L/a_s}\big)^2 \big|\vec{n}\big|^2 \Big
 / \Big( (a_t E_{\vec{n}})^2 - (a_t m)^2 \Big) \right]^{1/2}$, with the two volumes ($L/a_s=20,24$) and the different mesons offset slightly for clarity. The orange line and band indicate, respectively, the value and uncertainty on $\xi$ we use when investigating $\rho\pi$ scattering as described in the text.}
\label{ani_fig}
\end{figure}

The extracted energies are shown in Figure~\ref{ani_fig} along with the results of fits using Eq.~(\ref{disp_rel}).  For the $\rho$, the energies of the different helicity components were calculated independently from each relevant irrep of $\text{LG}(\vec{p})$, e.g. at $\vec{p}=[001]$ the $\lambda=0$ energies were calculated from the $[001]A_1$ irrep and $|\lambda|=1$ from $[001]E_2$. From the figure, it can be seen that the $\xi$ values extracted from the $\pi$ and the $|\lambda|=1$ $\rho$ are in reasonable agreement, but the value from the $\lambda = 0$ $\rho$ differs from the $\pi$ at the 2\% level\footnote{The energy splitting between different helicity components of the vector can be seen for calculations on a $16^3 \times 128$ lattice with the same lattice action in previous works -- see Figures 12 and 13 in Ref.~\cite{Thomas:2011rh} and Figure 4 in Ref.~\cite{Shultz:2015pfa}.}. This discrepancy is dominated by discretisation effects and we propagate a conservative estimate of systematic uncertainty by using a value of $\xi = 3.486(43)$, derived by considering the smallest and largest values within one standard deviation of the mean from the fits in Figure \ref{ani_fig}. Because the \mbox{meson-meson} interactions in $I=2$ $\rho\pi$ scattering are weak and the corresponding energy shifts small, the uncertainty on $\xi$ is found to be the largest source of systematic uncertainty on the scattering amplitudes.

\section{Finite-volume spectra for $\rho\pi$ in isospin-2 \label{Sec:FV}}
\begin{figure}[tb]
\vspace{-1cm}
    \centering
    \includegraphics[scale=0.575]{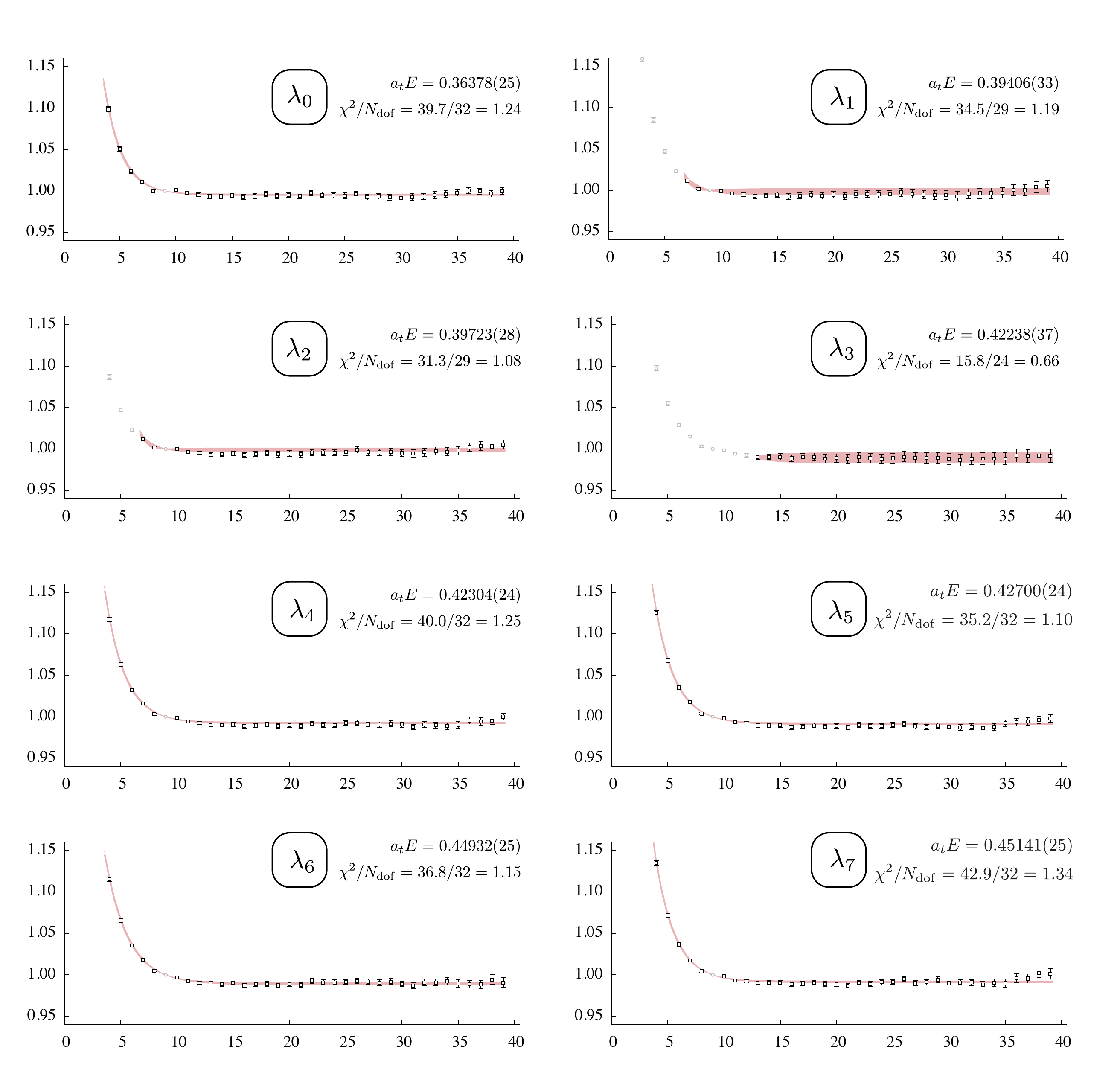}
\caption{Principal correlators, $\lambda_n(t,t_0=9)$, plotted as $e^{E_n(t-t_0)}\lambda_n(t,t_0)$,  from a variational analysis of the $[000]T_1^+$ irrep on the lattice with $L/a_s=24$. Curves show the results of fits as described in Section \ref{Sec:Spectrum_Determination}.}
	\label{princorrplot}
\end{figure}

\begin{figure}[tb]
    \centering
    \includegraphics[scale=2]{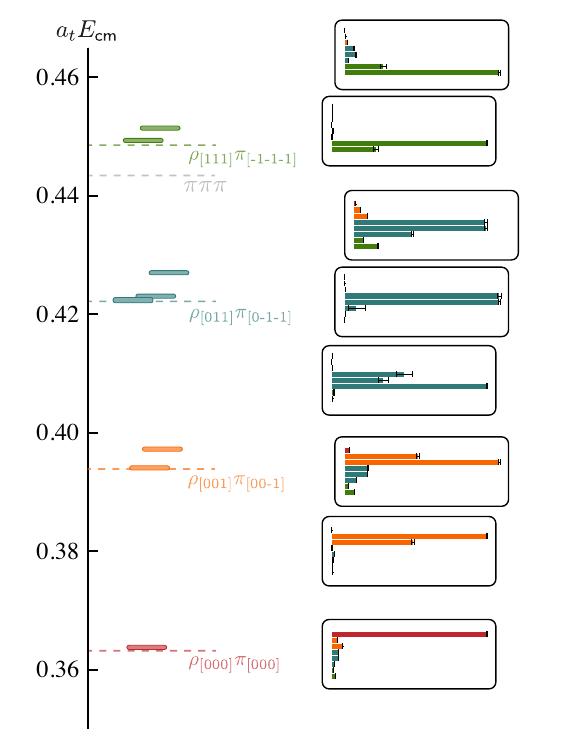}
    
\caption{Left: finite-volume energy levels in the $[000]T_1^+$ irrep on the lattice with ${L/a_s=24}$.  Dashed lines indicate the location of \mbox{non-interacting} energies.  Right: histograms showing the corresponding operator-state overlaps, $Z_i^n=\braket{n|\mathcal{O}^\dagger_i(0)|0}$, for the operators ordered as in Table \ref{i2_op_tab}. The colours reflect the \mbox{non-interacting} energies associated with each operator. The overlaps are normalised such that the largest value for any given operator across all energy levels is equal to one.}
	\label{histos}
\end{figure}

To determine finite-volume energy spectra for $I=2$ $\rho \pi$, matrices of correlation functions were calculated, using bases of \mbox{meson-meson} operators as outlined in Section \ref{Sec:Spectrum_Determination}, for all irreps $\vec{P}\;\Lambda$ where $\big|\vec{P}\big|^2 \leq 4\big(\frac{2\pi}{L}\big)^2$. Table \ref{i2_op_tab} shows the operators used in the $T_1^+$ irrep at rest and the $A_2$ irreps \mbox{in-flight} (operator lists for the other irreps are shown in Tables~\ref{optabB1} - \ref{optabB4}) -- note the multiple linearly-independent operators appearing at many of the non-interacting energies as discussed in Section~\ref{tw_me_op}. For each irrep, the finite-volume spectrum was extracted by applying the variational method as discussed in Section \ref{Sec:Spectrum_Determination}. As an example, we show the lowest eight principal correlators for the $T_1^+$ irrep in Figure~\ref{princorrplot}, and in Figure~\ref{histos} we present the corresponding spectrum and \mbox{operator-state} matrix elements, $Z_i^n\equiv\braket{n|\mathcal{O}^\dagger_i(0)|0}$. Figure~\ref{histos} shows that the matrix of correlation functions is nearly block diagonal in the momentum-based operator construction with respect to operators with the same $E_{\text{n.i}}$, and that different linear combinations of the multiple \mbox{meson-meson} operators, corresponding to the same $E_{\text{n.i}}$, are distinguishing the $N_{\text{lin}}$ nearly degenerate energy levels.

\begin{table}[tb]
\small
\begin{tabular}{r : r : r : r : r}
  \multicolumn{1}{c:}{$[000]T^+_1$} 
  &  \multicolumn{1}{c:}{$[001]A_2$ } 
  &  \multicolumn{1}{c:}{$[011]A_2$} 
  &  \multicolumn{1}{c:}{$[111]A_2$} 
  &  \multicolumn{1}{c}{$[002]A_2$} \\[0.5ex]
\hline
 $\rho_{[000]}\pi_{[000]}$ & $\rho_{[001]}\pi_{[000]}$ & $\rho_{[011]}\pi_{[000]}$ & $\rho_{[111]}\pi_{[000]}$ & $\rho_{[001]}\pi_{[001]}$ \\[0.5ex] 
 $\{2\} \;\rho_{[001]}\pi_{[00\text{-}1]}$ & $\rho_{[000]}\pi_{[001]}$ & $\{2\}\;\rho_{[001]}\pi_{[010]}$  & $\{2\}\;\rho_{[011]}\pi_{[100]}$ & $\rho_{[002]}\pi_{[000]}$ \\[0.5ex]
 $\{3\}\;\rho_{[011]}\pi_{[0\text{-}1\text{-}1]}$ & $\{2\}\;\rho_{[011]}\pi_{[0\text{-}10]}$ & $\rho_{[000]}\pi_{[011]}$  & $\{2\}\;\rho_{[100]}\pi_{[011]}$ & $\{2\}\;\rho_{[011]}\pi_{[0\text{-}11]}$ \\[0.5ex]
 $\{2\}\;\rho_{[111]}\pi_{[\text{-}1\text{-}1\text{-}1]}$ & $\{2\}\;\rho_{[010]}\pi_{[0\text{-}11]}$ & $\{2\}\;\rho_{[111]}\pi_{[\text{-}100]}$  & $\rho_{[000]}\pi_{[111]}$ & $\rho_{[000]}\pi_{[002]}$ \\[0.5ex] 
  & $\rho_{[002]}\pi_{[00\text{-}1]}$ & $\{3\}\rho_{[110]}\pi_{[\text{-}101]}$  & {\color{gray}$\mathit{ \{2\}\;\rho_{[112]}\pi_{[00\text{-}1]} }$} &  {\color{gray}$\mathit{ \{2\}\;\rho_{[012]}\pi_{[0\text{-}10]} }$} \\[0.5ex] 
  & $\{2\}\; \rho_{[111]}\pi_{[\text{-}1\text{-}10]}$ & $\{2\}\;\rho_{[100]}\pi_{[\text{-}111]}$ & {\color{gray}$\mathit{ \{3\}\;\rho_{[012]}\pi_{[10\text{-}1]} }$} & $\{2\}\;\rho_{[111]}\pi_{[\text{-}1\text{-}11]}$ \\[0.5ex]
  & $\{2\}\;\rho_{[110]}\pi_{[\text{-}1\text{-}11]}$ & {\color{gray}$\mathit{ \{2\}\;\rho_{[012]}\pi_{[00\text{-}1]} }$} & $\{2\}\;\rho_{[002]}\pi_{[11\text{-}1]}$  & {\color{gray}$ \mathit{ \{2\}\;\rho_{[010]}\pi_{[0\text{-}12]} }$} \\[0.5ex]
  & $\rho_{[00\text{-}1]}\pi_{[002]}$ & $\{2\}\;\rho_{[002]}\pi_{[01\text{-}1]}$ &  $\{2\}\; \rho_{[11\text{-}1]}\pi_{[002]}$ & {\color{gray}$\mathit{ \{2\}\;\rho_{[112]}\pi_{[\text{-}1\text{-}10]} }$} \\[0.5ex]
  & {\color{gray}$\mathit{ \{2\}\;\rho_{[012]}\pi_{[0\text{-}1\text{-}1]} }$} & $\{2\}\; \rho_{[01\text{-}1]}\pi_{[002]}$ & {\color{gray}$ \mathit{ \{3\}\;\rho_{[01\text{-}1]}\pi_{[102]} }$} & {\color{gray}$\mathit {\{2\}\; \rho_{[\text{-}1\text{-}10]}\pi_{[112]}}$}\\[0.5ex]
  & & {\color{gray}$\mathit{ \{2\}\;\rho_{[00\text{-}1]}\pi_{[012]} }$} & {\color{gray}$\mathit{  \{2\}\; \rho_{[00\text{-}1]}\pi_{[112]} }$} & \\[0.5ex]
  & & {\color{gray}$\mathit{ \{3\}\;\rho_{[112]}\pi_{[\text{-}10\text{-}1]} }$} & & \\[0.5ex]
\hline
$8\,\text{ops.}$ & $12\,\text{ops.}$ & $15\,\text{ops.}$ & $10\,\text{ops.}$ & $7\,\text{ops.}$ \\
\end{tabular}
\caption{Meson-meson operators in the $\bm{27}$ of $\text{SU}(3)_{\text{F}}$ flavour, ordered by increasing \mbox{non-interacting} energy (see Section~\ref{tw_me_op}), for various irreps $\vec{P}\;\Lambda$. The operators, $\rho_{\vec{p}_1}\pi_{\vec{p}_2}$, are constructed from optimised $\rho$ and $\pi$ operators with momentum types $\vec{p}_1$ and $\vec{p}_2$ respectively; different momentum directions are summed over as in Eq.~(\ref{MM}). $\{N_{\text{lin}}\}$ denotes the number of linearly-independent \mbox{meson-meson} operators at the corresponding non-interacting energy when there is more than one. All operators with corresponding \mbox{non-interacting} energies $a_tE_\mathsf{cm} \leq 0.455$ for $L/a_s = 24$ are displayed. Those in grey italic were not included in the operator basis.}
\label{i2_op_tab}
\end{table}

\begin{figure}[tb]  
\includegraphics[scale=1]{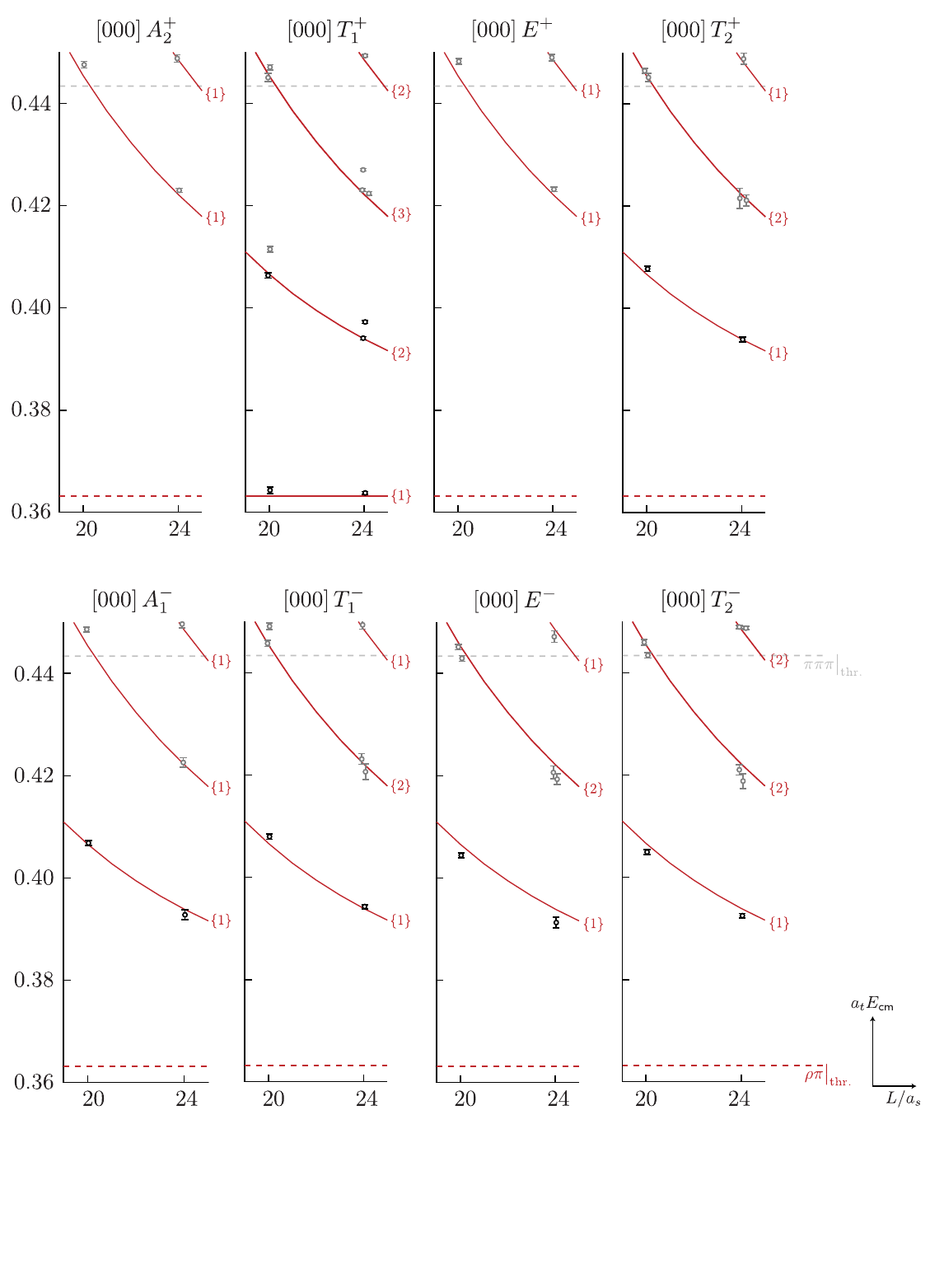}
\centering
\vspace{-2cm}
\caption{Energy spectra in irreps at rest. Black and grey points, slightly displaced in $L/a_s$ for clarity, show the extracted energy levels with statistical uncertainties below and above $a_tE_\mathsf{cm}=0.41$ respectively. Points in grey are not used in the subsequent analysis in Section~\ref{Sec:Results}. Dashed lines show the $\rho\pi$ and $\pi\pi\pi$ thresholds. Solid red curves indicate \mbox{non-interacting} \mbox{meson-meson} energies, labelled with their degeneracies.}
\label{vm_p000}
\end{figure}

\begin{figure}[tb]  
	\includegraphics[scale=1]{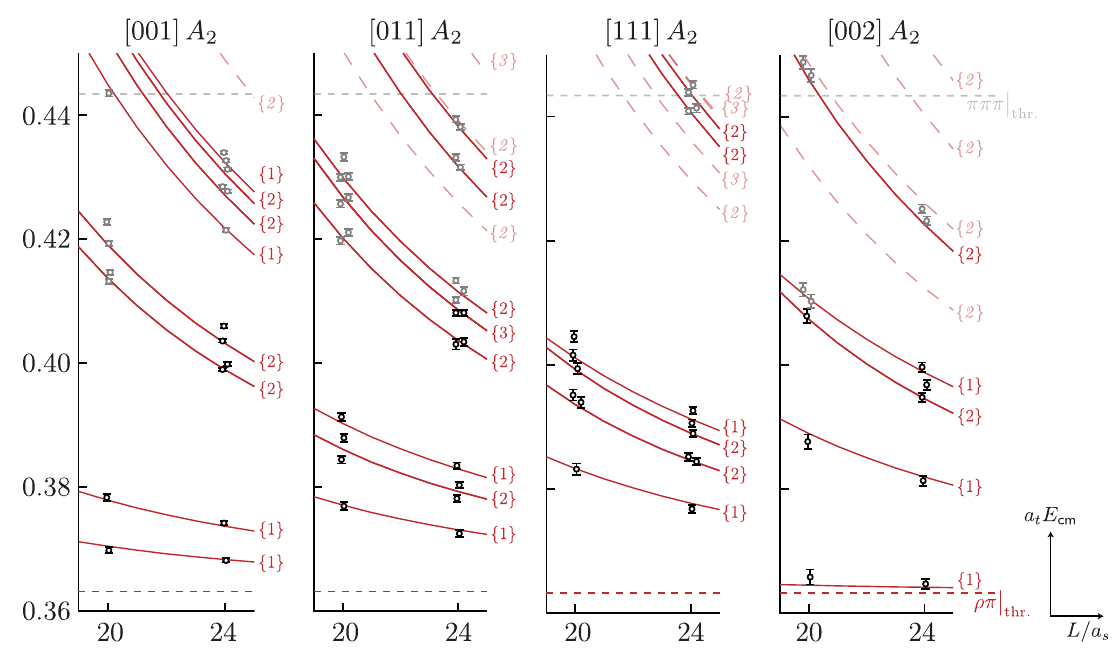}
	\centering
	\caption{As Figure \ref{vm_p000} but for $A_2$ irreps with $\vec{P}\neq \vec{0}$. Dashed red curves indicate \mbox{non-interacting} \mbox{meson-meson} energies corresponding to operators not included in the basis. Errors on the points show the statistical uncertainty added in quadrature to the systematic uncertainty from the uncertainty placed on $\xi$.}
\label{vm_DA2}
\end{figure}

In Figures~\ref{vm_p000} and~\ref{vm_DA2} we show the volume dependence of the extracted energies for all irreps at rest and $A_2$ irreps \mbox{in-flight}. Spectra for other \mbox{in-flight} irreps can be found in Figure~\ref{vm_app} in Appendix~\ref{App:Finite_Volume_Spectra}. The energy levels used in the scattering analysis are included as supplementary material. Figure~\ref{vm_DA2} illustrates the dense distribution of energy levels typical of \mbox{in-flight} irreps, a consequence of the reduced symmetry, and the multiple energy levels which would be degenerate in the absence of interactions. Nevertheless, it can be seen that all the energy levels can be extracted with good statistical precision. Since we choose to restrict our operator bases to include only \mbox{single-meson} operators with momentum $|\vec{n}|^2 \le 4$, we will only extract scattering amplitudes for $a_t E_\mathsf{cm}\leq 0.41$, below the \mbox{non-interacting} energy corresponding to the lowest excluded operator\footnote{The lowest-lying excluded operator, across all irreps and volumes, is $\rho_{[012]}\pi_{[0\text{-1}0]}$, which corresponds to a  \mbox{non-interacting} energy of $a_tE_\mathsf{cm}=0.4124$ on the lattice with $L/a_s = 24$.}. No other \mbox{meson-meson} scattering channels have thresholds below the $\pi\pi\pi$ threshold which opens at $a_tE_\mathsf{cm}=0.443$.

Some qualitative expectations for the behaviour of scattering amplitudes can be inferred from the spectra presented in Figures~\ref{vm_p000} and~\ref{vm_DA2}. There are clearly no large departures from the \mbox{non-interacting} spectra, the number of energy levels is the same as the number expected in the absence of interactions, and no energy levels lie systematically below the $\rho\pi$ threshold. These observations likely indicate the absence of narrow resonances or bound-states, and suggest that only a relatively weak interaction is present. In order to get a quantitative understanding we proceed to analyse the spectra using the quantisation condition discussed in Section~\ref{secFV}.


\section{Scattering amplitudes for $\rho\pi$ in isospin-2 \label{Sec:Results}}
The relationship between infinite-volume scattering amplitudes and finite-volume energy levels, originally developed by L\"{u}scher \cite{Luscher:1985dn,Luscher:1986pf,Luscher:1990ux,Luscher:1991cf}, has been extended by numerous works \cite{Rummukainen:1995vs,He:2005ey,Christ:2005gi,Kim:2005gf,Guo:2012hv,Hansen:2012tf,Briceno:2012yi,Briceno:2014oea,Gockeler:2012yj} to incorporate the most general two-body scattering processes. We summarised the essence of this quantisation condition in Eq.~(\ref{full_luescher}) and now discuss it in more detail.

For a single vector-pseudoscalar scattering channel in a cubic spatial box with periodic boundary conditions, the quantisation condition for the spectrum in irrep $\Lambda$ at momentum $\vec{P}=\frac{2\pi}{L}\vec{n}$ can be written as,
\begin{equation}\label{QCBSUB}
	\det\bigg[
	\delta_{\ell \ell''} \, \delta_{JJ''} \, \delta_{nn''} + i \rho \, t_{\ell J n, \ell'J'n'} \, 
	\bigg(
	\delta_{\ell' \ell''} \, \delta_{J'J''} \, \delta_{n'n''} + i \, \overline{\mathcal{M}}^{\, \vec{n},\, \Lambda}_{\ell' J'n',\, \ell''J''n''}
	\bigg)
	\bigg]=0 \, .
\end{equation}
In this expression the determinant is evaluated over matrices whose rows and columns are labelled by $(\ell, J, n)$, for partial-waves $\threelJ$ subduced into the irrep $\Lambda$, where $n$ denotes the $n^{th}$ \emph{embedding} of the partial-wave\footnote{For example, in the $[011]\, A_2$ irrep there are two embeddings of the partial-wave $\threePtwo$ -- see Table \ref{tab011}.}. The infinite-volume $\mathbf{t}$-matrix, with elements $t_{\ell Jn,\ell'J'n'}$, is diagonal in $J$ and $n$ but not in $\ell$ as discussed in Section~\ref{infvolsec}. The phase-space, $\rho=2k_\mathsf{cm}/E_\mathsf{cm}$, is a function of the centre-of-momentum frame momentum, ${k_\mathsf{cm}=\tfrac{1}{2E_\mathsf{cm}}\big(E_\mathsf{cm}^2-(m_\pi+m_\rho)^2\big)^{1/2}  \big(E_\mathsf{cm}^2-(m_\pi-m_\rho)^2\big)^{1/2}}$. The matrix $\overline{\boldsymbol{\mathcal{M}}}$ is a matrix of known functions of $E_\mathsf{cm}$ and $L$ that incorporates the effects of the finite volume; the explicit form of $\overline{\boldsymbol{\mathcal{M}}}$ and further details of the quantisation condition shown in Eq.~(\ref{QCBSUB}) are given in Appendix \ref{App:Details_Of_Luescher}.

In the case that no partial-waves are coupled dynamically, the $\mathbf{t}$-matrix is diagonal in $\ell$ and infinite-volume scattering in each partial-wave, $\threelJ$, can be described by a single real-valued energy-dependent parameter called the \mbox{phase-shift}, $\delta_{\threelJ}(E_\mathsf{cm})$. This appears in the scattering $\mathbf{t}$-matrix as $t_{\ell J n, \ell Jn}=\frac{1}{\rho}\exp[i\,\delta_{\threelJ}]\sin (\delta_{\threelJ})$. In an irrep where just a single partial-wave makes a non-negligible contribution to the quantisation condition, Eq.~(\ref{QCBSUB}) reduces to the form shown in Eq.~(\ref{elas_luescher}) -- this can be evaluated to give a \mbox{phase-shift} point, $\delta_{\threelJ}\big(  E^{(k)}_\mathsf{cm} \big)$, at each finite-volume energy level, $E^{(k)}_\mathsf{cm}$. 

Formally, the infinite number of partial-waves which subduce into the irrep $\Lambda$ appear in the quantisation condition. Even though the angular-momentum barrier suppresses the contributions of partial-waves of higher $\ell$ at low energies, for vector-pseudoscalar scattering multiple partial-waves with the same threshold behaviour can appear in a single irrep. For example, the $\threePone$ and $\threePtwo$ partial-waves both appear in $[011]\, A_1$. This prevents the use of a one-to-one mapping between energy levels and \mbox{phase-shift} points of the type given in Eq.~(\ref{elas_luescher}).

Furthermore, when two partial-waves are dynamically coupled, the scattering $\mathbf{t}$-matrix is not diagonal in $\ell$ and is described by three real energy-dependent parameters\footnote{Given the constraints from unitarity of the $\mathbf{S}$-matrix and the time-reversal symmetry of QCD.}. These can be expressed as two \mbox{phase-shifts} and an angle, as in Eq.~(\ref{tinf}). In this case, again, there is no one-to-one mapping between energy levels and \mbox{phase-shift} points.

One approach to determine scattering information when the energy spectrum is dependent on more than a single energy-dependent scattering parameter is to, as in Refs.~\cite{Dudek:2012gj,Moir:2016srx,Briceno:2017qmb,Dudek:2014qha,Wilson:2014cna,Dudek:2016cru,Wilson:2015dqa,Andersen:2017una}, \emph{parameterise the energy-dependence} of the $\mathbf{t}$-matrix. In this way, for any given set of parameter values, a finite-volume spectrum is predicted in each irrep by solving Eq.~(\ref{QCBSUB}). We follow the approach of Ref.~\cite{Wilson:2014cna} where this predicted spectrum is compared to the computed lattice spectrum using an appropriate $\chi^2$, as defined in Eq.~9 of \cite{Dudek:2012xn}, where correlations between energy levels on the same lattice volume are accounted for using the data covariance matrix. By minimising the $\chi^2$ with respect to the free parameters, the best description of the spectrum may be obtained. The sensitivity to the choice of scattering-amplitude parameterisation can be tested by using a variety of different parameterisations.

In the case of a single partial-wave not dynamically coupled to any others, a convenient parameterisation of,
\[
{t_{\ell Jn, \ell Jn}=\frac{1}{\rho}\exp \big[i\,\delta_{\threelJ}\big]  \sin (\delta_{\threelJ})  = \frac{E_\mathsf{cm}/2}{k_\mathsf{cm} \cot (\delta_{\threelJ}) - i k_\mathsf{cm}}   } \, ,
\] 
is the effective range expansion, 
\begin{equation}\label{eff_range}
k_\mathsf{cm}^{2\ell+1}\, \cot (\delta_{\threelJ}) = \frac{1}{a({\threelJ}|{\threelJ})} + \frac{1}{2} \, r({\threelJ}|{\threelJ}) \, k_\mathsf{cm}^2 + \mathcal{O}\big(k_\mathsf{cm}^4 \big) \, ,
\end{equation}
where the constants $a({\threelJ}|{\threelJ})$ and $r({\threelJ}|{\threelJ})$ are respectively the scattering length and effective range of the partial-wave $\threelJ$, and the threshold behaviour of the amplitude, controlled by the value of $\ell$, is explicitly included by construction.
 
For partial-waves of equal $J$ but different $\ell$ that can couple dynamically, the $\mathbf{K}$-matrix formalism is a useful way of expressing the unitarity of the $\mathbf{S}$-matrix in terms of a real symmetric matrix, $\mathbf{K}(s)$.\footnote{Previous lattice QCD calculations \cite{Wilson:2014cna,Wilson:2015dqa,Dudek:2016cru} have demonstrated the effectiveness of the $\mathbf{K}$-matrix formalism in describing many resonant and non-resonant features of coupled-channel scattering.}~The inverse of the $\mathbf{K}$-matrix is related to the inverse of the $\mathbf{t}$-matrix by,
\begin{equation}\label{ktinv}
\big[t^{-1}(s)\big]_{\ell J, \ell' J} = 
\frac{1}{(2k_\mathsf{cm})^\ell}  \, 
\big[K^{-1}(s)\big]_{\ell J, \ell' J} \,
\frac{1}{(2k_\mathsf{cm})^{\ell'}} 
+ \delta_{\ell \ell'}\, I(s) \,,
\end{equation}
where $s=E_\mathsf{cm}^2$.\footnote{In the quantisation condition, should multiple embeddings of $J$ appear, the $\mathbf{t}$-matrix is repeated $n$ times in a block diagonal form.} The powers of $k_\mathsf{cm}$ ensure the correct behaviour at threshold. Unitarity of the $\mathbf{S}$-matrix is guaranteed provided that $\text{Im}\, I(s)=-\rho(s)$ for energies above the vector-pseudoscalar threshold and $\text{Im}\, I(s)=0$ below threshold. The real part of $I(s)$ is arbitrary, with the simplest choice being $\text{Re}\,  I(s)=0$. An alternative which improves the analytic properties of the amplitude, known as the Chew-Mandelstam prescription~\cite{Chew:1960iv}, constructs $\text{Re}\,I(s)$ using a dispersive integral of $\rho(s)$. The implementation of this prescription used here mirrors that in Ref.~\cite{Wilson:2014cna} and we choose to subtract such that $\text{Re}\,I(s)=0$ at threshold. Hereinafter, we use this prescription unless otherwise specified.

The $\mathbf{K}$-matrix can be generalised to handle the case relevant to the finite volume where different $J$ values, which are uncoupled in an infinite volume, become coupled in the determinant of Eq.~(\ref{QCBSUB}). This is achieved by forming a block-diagonal matrix out of the $\mathbf{K}$-matrices for each $J$. For example, the $\mathbf{t}$-matrix described in Eq.~(\ref{tinffin}) will feature the $\mathbf{K}$-matrix,
\begin{equation}
\mathbf{K}=
\begin{bmatrix}
 K(\threeSone |\!\threeSone)(s) & K(\threeSone |\!\threeDone)(s) & 0 \\
 K(\threeSone |\!\threeDone)(s) & K(\threeDone |\!\threeDone)(s) & 0 \\
 0 & 0 & K(\threeDthree |\!\threeDthree)(s)
\end{bmatrix}
\end{equation}
where $ K({\threelJ} | {^3\ell'_{J'}})(s) \equiv K_{\ell J,\ell' J'}(s)$ is a real function of $s$. 

A simple choice of parameterisation for the $\mathbf{K}$-matrix is to express each element as a finite-order polynomial in $s$,
\begin{equation}\label{Kparams}
K_{\ell J, \ell' J}(s) = \sum_{n\geq 0}^{N(\threelJ | \!\threelprimeJ)}   c_n(\threelJ | \!\threelprimeJ)\, s^n,
\end{equation}
where the coefficients $c_n(\threelJ | \!\threelprimeJ)$ are real parameters.

\subsection{Uncoupled $P$-wave scattering}\label{uncoupled-pwave}
\begin{figure}[tb]     
	\centering
	\noindent
	\makebox[\textwidth]{\includegraphics[scale=0.45]{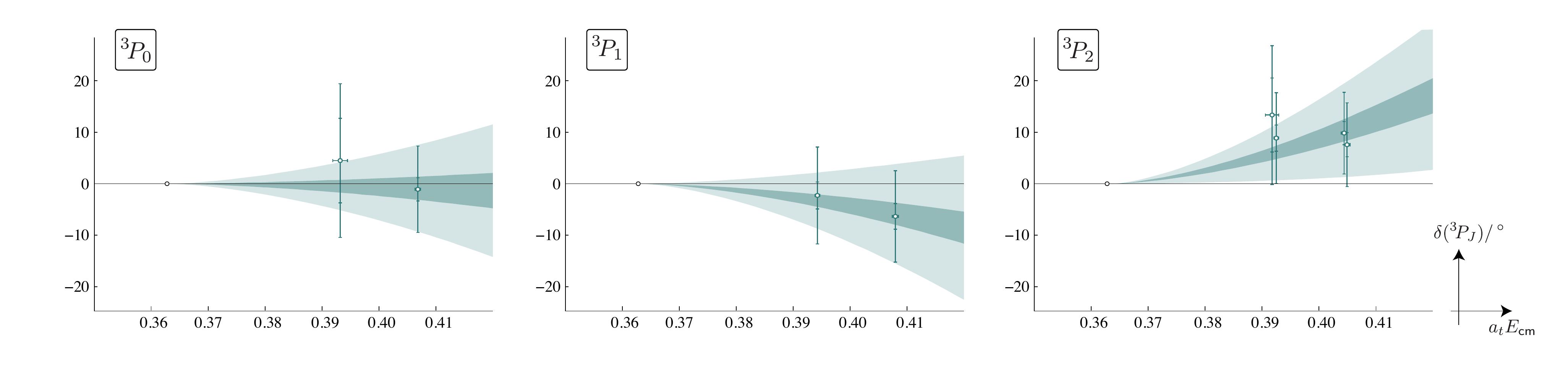}}
	\vspace{-10mm}
	\caption{Phase-shifts for the $\threePzero$, $\threePone$ and $\threePtwo$ partial-waves. The points are as described in the text. Inner bands reflect the statistical uncertainties on the \mbox{phase-shifts} from the fits ~(\ref{scat_lengths}) and outer bands reflect the combined statistical and systematic uncertainties.}
\label{pelas}
\end{figure}

As discussed above, when only a single partial-wave makes a non-negligible contribution to Eq.~(\ref{QCBSUB}), the finite-volume quantisation condition reduces to a one-to-one mapping from finite-volume energy levels to \mbox{phase-shift} values at those energies. For $I=2$ $\rho\pi$ scattering, we initially assume that the $\threePzero$, $\threePone$, $\threePtwo$ partial-waves dominate respectively the $[000]\,A_1^-$, $T_1^-$, $(E^-,T_2^-)$ irreps at low energy, proposing that the $F$-wave contributions can be neglected (see Table~\ref{tab000} for the partial-waves subduced into these irreps). Using the energy levels presented in Figure~\ref{vm_p000}, we obtain two \mbox{phase-shift} points from each irrep. These are shown in Figure~\ref{pelas} where the inner error bars show the statistical uncertainty on $E_\mathsf{cm}$ and $\delta_{{^3P_J}}(E_\mathsf{cm})$, while the outer error bars on $\delta_{{^3P_J}}(E_\mathsf{cm})$ also include a conservative estimate of the systematic error which was obtained by varying the hadron masses and, more importantly, the anisotropy within their uncertainties. We find the largest systematic variations occur when $a_tm_{\rho}$, $a_tm_{\pi}$ are large and $\xi$ is small, and vice-versa\footnote{For $a_tm_{\rho}$, $a_tm_{\pi}$ small and $\xi$ large we find a compatible order of magnitude of variation in the parameters but of opposite sign. We therefore quote the systematic error as symmetric about the mean.}, consistent with the observation that this causes the largest changes in the \mbox{non-interacting} energies, $E_\mathsf{n.i.}$.

To interpolate the scattering amplitudes in the energy range being considered, we parameterise the energy dependence of the $\mathbf{t}$-matrix using an effective range expansion, Eq.~(\ref{eff_range}), truncated at the scattering length, ${k_\mathsf{cm}^{2\ell+1}}\, \cot (\delta_{\threelJ}) = a({\threelJ}|{\threelJ})^{-1}$, and minimise a $\chi^2$ with respect to $a({\threelJ}|{\threelJ})$. We fit independently for each partial-wave obtaining,
\begin{align}\label{scat_lengths}
a(\threePzero|\!\threePzero)=& \,(-21 \pm 53 \pm 145) \cdot a_t^3 & \chi^2/N_{\text{dof}} = 0.37/(2-1) = 0.37 \nonumber \\
a(\threePone|\!\threePone)=& \,(-133 \pm 49 \pm 172) \cdot a_t^3 & \chi^2/N_{\text{dof}} = 0.20/(2-1) = 0.20 \nonumber \\
a(\threePtwo|\!\threePtwo)=& \,(+273 \pm 58 \pm 184) \cdot a_t^3 & \chi^2/N_{\text{dof}} = 6.57/(4-1) = 2.19 &,
\end{align}
where again the first error reflects the statistical uncertainty and the second error is an estimate of the systematic uncertainty.

The energy dependencies of the \mbox{phase-shifts} corresponding to these scattering-length descriptions are displayed in Figure \ref{pelas}. It is clear that the systematic uncertainties are dominating the uncertainties -- this is a consequence of the relatively large uncertainty assigned to $\xi$,\footnote{because of the slightly different $\xi$ obtained from the helicity $0$ and $\pm1$ components of the $\rho$} coupled with the rather weak interaction in this scattering channel which leads to small shifts of energies from their \mbox{non-interacting} values.
\subsection{$S,P,D$-wave scattering including dynamically-coupled partial-waves}\label{global_analysis}
In general, irreps feature a number of partial-waves and so there is not a one-to-one mapping between energy levels and scattering amplitudes. To use the information from the energy levels in all the irreps, we perform a \emph{global} analysis of the finite-volume spectra presented in Figures~\ref{vm_p000}, \ref{vm_DA2} and \ref{vm_app}: each energy level provides a constraint on a combination of partial-wave amplitudes at that energy. To do this, as described above, we parameterise the energy-dependence of the block-diagonal $\mathbf{t}$-matrix and vary the parameters to give the best description of the finite-volume spectra. We allow for non-negligible $\rho\pi$ isospin-2 amplitudes in the $\threeSone$, $\threePzero$, $\threePone$, $\threePtwo$, $\threeDone$, $\threeDtwo$ and $\threeDthree$ partial-waves, including the dynamical couplings between the $\threeSone$ and $\threeDone$ waves and the $\threePtwo$ and $\threeFtwo$ waves.

A number of polynomial parameterisations of the $\mathbf{K}$-matrix were considered and one example giving a good description of the 141 energy levels below $a_tE_\mathsf{cm}=0.41$ is provided by the fit shown in Table \ref{ALLparams} where a $\mathbf{K}$-matrix parameterisation with 11 parameters was used: there are linear plus constant terms in $K(\threeSone|\!\threeSone)$, $K(\threePone|\!\threePone)$ and $K(\threePtwo|\!\threePtwo)$, and constant terms for all other relevant $K({\threelJ}|{\threelprimeJ})$ except $K({\threePtwo}|{\threeFtwo})=0$.
The table also gives statistical uncertainties, estimates of systematic uncertainties from varying $a_tm_\pi$, $a_tm_\rho$ and $\xi$, and correlations between the parameters.  We refer to this parameterisation and set of fit values as our reference amplitude.
\begin{table}[tb]
	\noindent\resizebox{\linewidth}{!}{
  \centering
  $
  \begin{array}{l l}
     c_0({\threeSone}|{\threeSone}) &= -1.61 \pm 0.07 \pm 0.79 \\
     c_1({\threeSone}|{\threeSone})  &= (4.75 \pm 0.44 \pm 5.37) \cdot a_t^2 \\
     c_0({\threeSone}|{\threeDone}) &= (-5.28 \pm 0.55 \pm 0.51)\cdot a_t^2 \\
     c_0({\threePzero}|{\threePzero})  &= (-5.98 \pm 0.61 \pm 4.70) \cdot a_t^2 \\
     c_0({\threePone}|{\threePone})  &= (-33.6 \pm 1.7 \pm 17.7)\cdot a_t^2 \\
     c_1({\threePone}|{\threePone})  &= (150 \pm 11 \pm 128)\cdot a_t^4  \\
     c_0({\threePtwo}|{\threePtwo})  &= (83.4 \pm 1.5 \pm 40.7) \cdot a_t^2 \\
     c_1({\threePtwo}|{\threePtwo})  &= (-459 \pm 9 \pm 277) \cdot a_t^4 \\
     c_0({\threeDone}|{\threeDone})  &= (-56 \pm 15 \pm 31) \cdot a_t^4 \\
     c_0({\threeDtwo}|{\threeDtwo})  &= (-102 \pm 12 \pm 60) \cdot a_t^4 \\
     c_0({\threeDthree}|{\threeDthree})  &= (-49 \pm 15 \pm 84) \cdot a_t^4 \\
  \end{array}
  \begin{array}{c}
    \left[
    \renewcommand{\arraystretch}{1.345}
    \scalemath{0.75}{
    \begin{array}{r r r r r r r r r r r r}
      1.00 & -0.98 & 0.04 & 0.11 & 0.02 & 0.02 & 0.03 & 0.02 & 0.06 & 0.08 & 0.10 \\
       & 1.00 & -0.11 & -0.05 & -0.01 & -0.01 & -0.02 & -0.01 & -0.05 & -0.01 & -0.05 \\
      & & 1.00 & 0.09 & 0.01 & 0.03 & 0.04 & 0.02 & 0.26 & -0.03 & 0.22 \\
      & & & 1.00 & 0.10 & 0.11 & 0.14 & 0.16 & 0.02 & 0.26 & 0.77 \\
      &  & &  & 1.00 & -0.95 & 0.04 & 0.01 & 0.02 & 0.06 & 0.08 \\
      &  & &  & & 1.00 & 0.01 & 0.04 & 0.01 & 0.04 & 0.10 \\
       &  &  & & & & 1.00 & -0.92 & 0.04 & 0.08 & 0.14 \\
       &  &  & & & & & 1.00 & 0.03 & 0.09 & 0.10 \\
      & & & & & & & & 1.00 & 0.46 & -0.09 \\
       &  &  & &  & && &  & 1.00 & 0.06 \\
      &  &  & & & &  & &  &  & 1.00 \\
    \end{array}
    } 
    \right] \\ 
  \end{array} 
  $ 
  \caption{A reference fit as described in the text with $\chi^2/N_{\text{dof}} = 1.42$. The first uncertainty in each case is statistical and the second is an estimate of the systematic uncertainty as described in the text. Correlations between the $\mathbf{K}$-matrix parameters are displayed on the right. Parameters not shown were fixed to zero.}
	\label{ALLparams}
}
\end{table}

\begin{figure}[tb]     
	\centering
	\vspace*{-8mm}
    \includegraphics[scale=1]{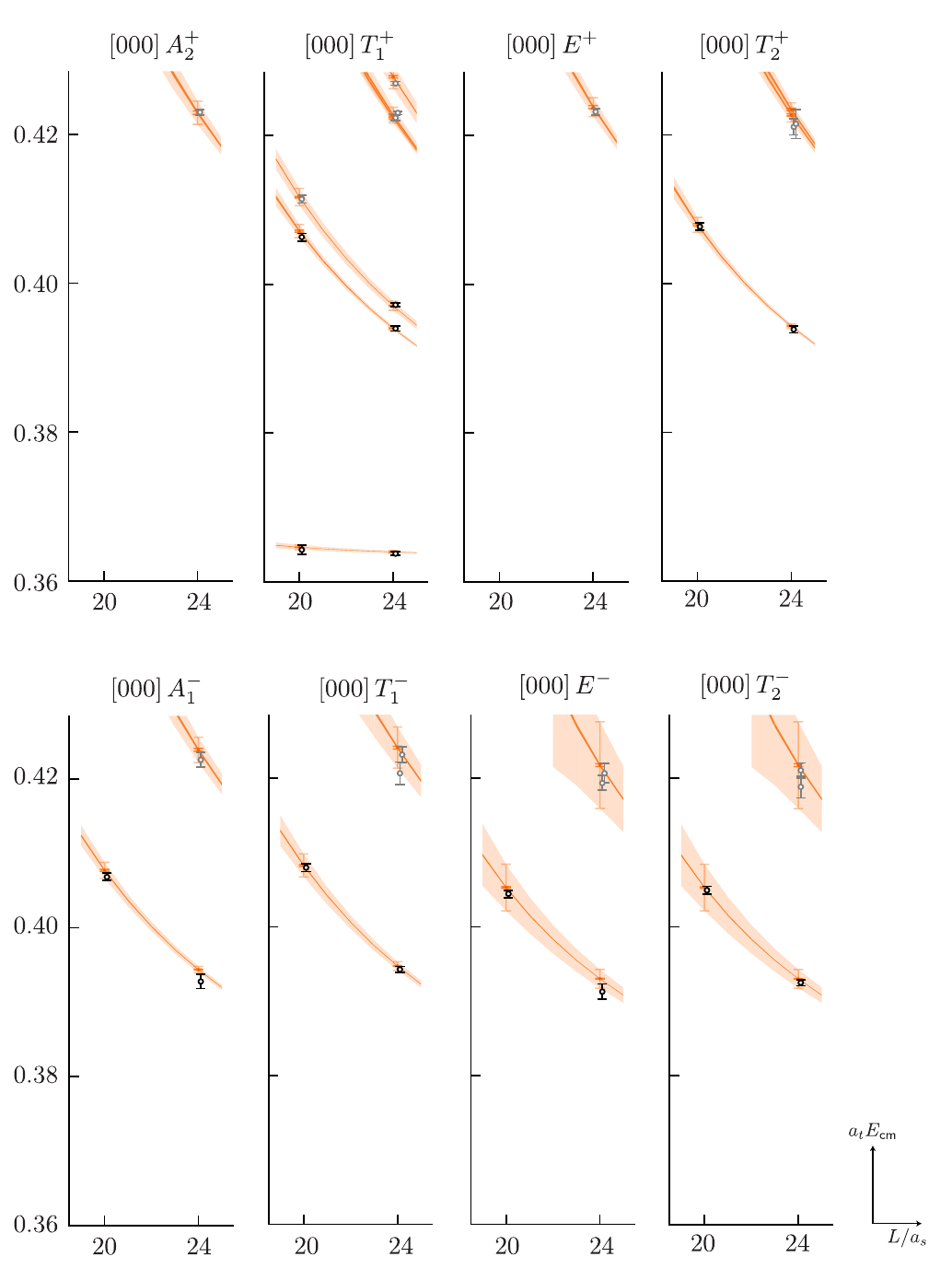}
	\caption{Volume-dependent spectra for irreps with $\vec{P}=\vec{0}$. Black and grey points, slightly displaced in $L/a_s$ for clarity, are, as in Figure \ref{vm_p000}, energy levels extracted from analyses of correlation functions. Orange points and bands show solutions to Eq.~(\ref{QCBSUB}) for the reference $\mathbf{K}$-matrix parameterisation in Table \ref{ALLparams}. The inner dark orange error bars/error bands reflect the statistical uncertainties and the outer lighter orange error bars/error bands also include systematic uncertainties.}
	\vspace*{-2mm}
	\label{fv_stat_sys_rest}
\end{figure}

\begin{figure}[tb]     
	\vspace*{-8mm}
	\centering
	\includegraphics[scale=1]{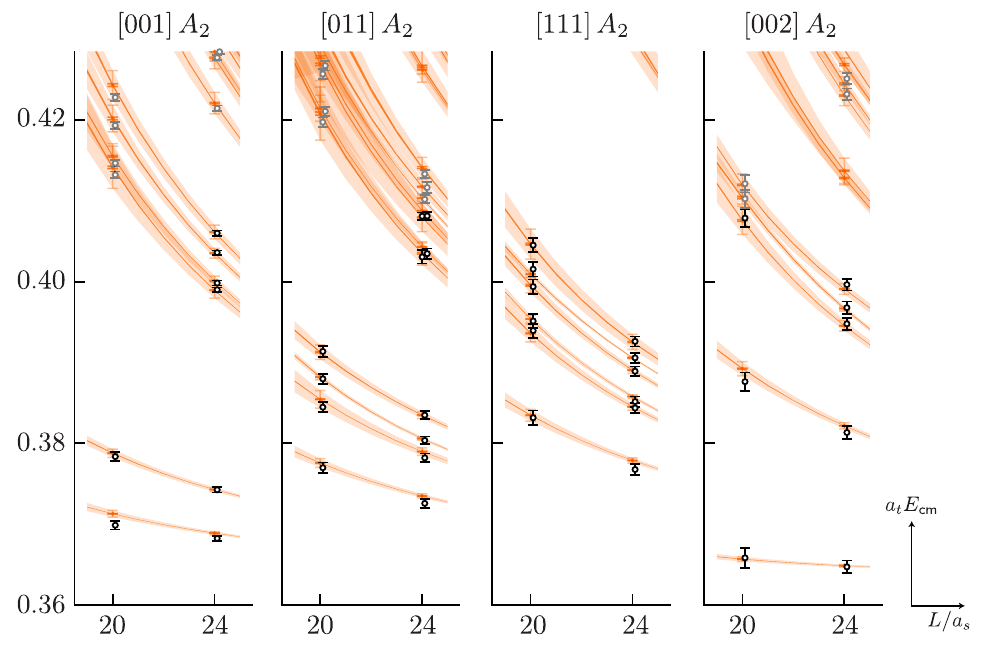}
	\caption{As Figure \ref{fv_stat_sys_rest} but for $A_2$ irreps with $\vec{P}\neq \vec{0}$.}
	\vspace*{-1mm}
\label{fv_stat_sys_A2}
\end{figure}

Presented in Figures \ref{fv_stat_sys_rest} and \ref{fv_stat_sys_A2} are the finite-volume spectra obtained by solving Eq.~(\ref{QCBSUB}) for the reference amplitude. The levels previously plotted in Figures~\ref{vm_p000} and~\ref{vm_DA2} are also shown on the figure and we observe very good agreement between the two sets of energy levels (as expected from the $\chi^2$). The reference amplitude successfully predicts the location of levels which were not used to constrain the parameterisation (grey points), but a couple of features should be noted. Firstly, in Figure~\ref{fv_stat_sys_rest} some levels are apparently missed by the scattering parameterisation in the the $E^-$, $T_1^-$ and $T_2^-$ irreps around $a_tE_\mathsf{cm}=0.42$. The presence of these levels relies upon the inclusion of $F$-wave scattering amplitudes, which are neglected in the reference amplitude. Secondly, in Figure~\ref{fv_stat_sys_A2} the $A_2$ irreps with $\vec{P}=[011]$ and $\vec{P}=[002]$ appear to have energy levels missing in the lattice QCD calculation around $a_tE_\mathsf{cm}=0.425$ and $a_tE_\mathsf{cm}=0.415$ respectively. This is expected because the corresponding vector-pseudoscalar operators were not included in the bases used (see Section \ref{Sec:FV}, Table \ref{i2_op_tab} and Figure~\ref{vm_DA2}).

A wide range of possible parameterisations that allow non-zero values for all constants $c_n(\threelJ|\!\threelprimeJ)$ provided $\ell+\ell'+2n \le 4$ were considered. This ensures the $\mathbf{K}$-matrix has parameter freedom in all terms up to order $a_t^4$.\footnote{Including terms with higher powers of $a_t$ did not significantly improve the quality of fit.}~Table~\ref{paramvar} in Appendix~\ref{App:Global_Fit_Parameterisations} shows a selection of these fits along with the corresponding $\chi^2/N_{\text{dof}}$. Parameterisations without freedom in the $K({\threeSone} | {\threeDone})(s)$ polynomial are not able to give a good description of the finite-volume spectra, a point we return to in Section~\ref{eps}. However, a $K({\threePtwo} | {\threeFtwo})(s)$ term does not appear to be required -- this is consistent with expectations that the dynamical mixing between $\threePtwo$ and $\threeFtwo$ is suppressed by the angular momentum barrier at these relatively low energies just above threshold.

$\mathbf{K}$-matrix parameterisations which include pole terms, efficient at describing resonant behaviour and bound states, did not give a good description of the finite-volume spectra and we do not include such parameterisations in Table \ref{paramvar}. This is consistent with our qualitative observations on the spectra in Section \ref{Sec:FV}.

\begin{figure}
	\vspace*{-5mm}
		\centering
		\includegraphics[scale=1.415]{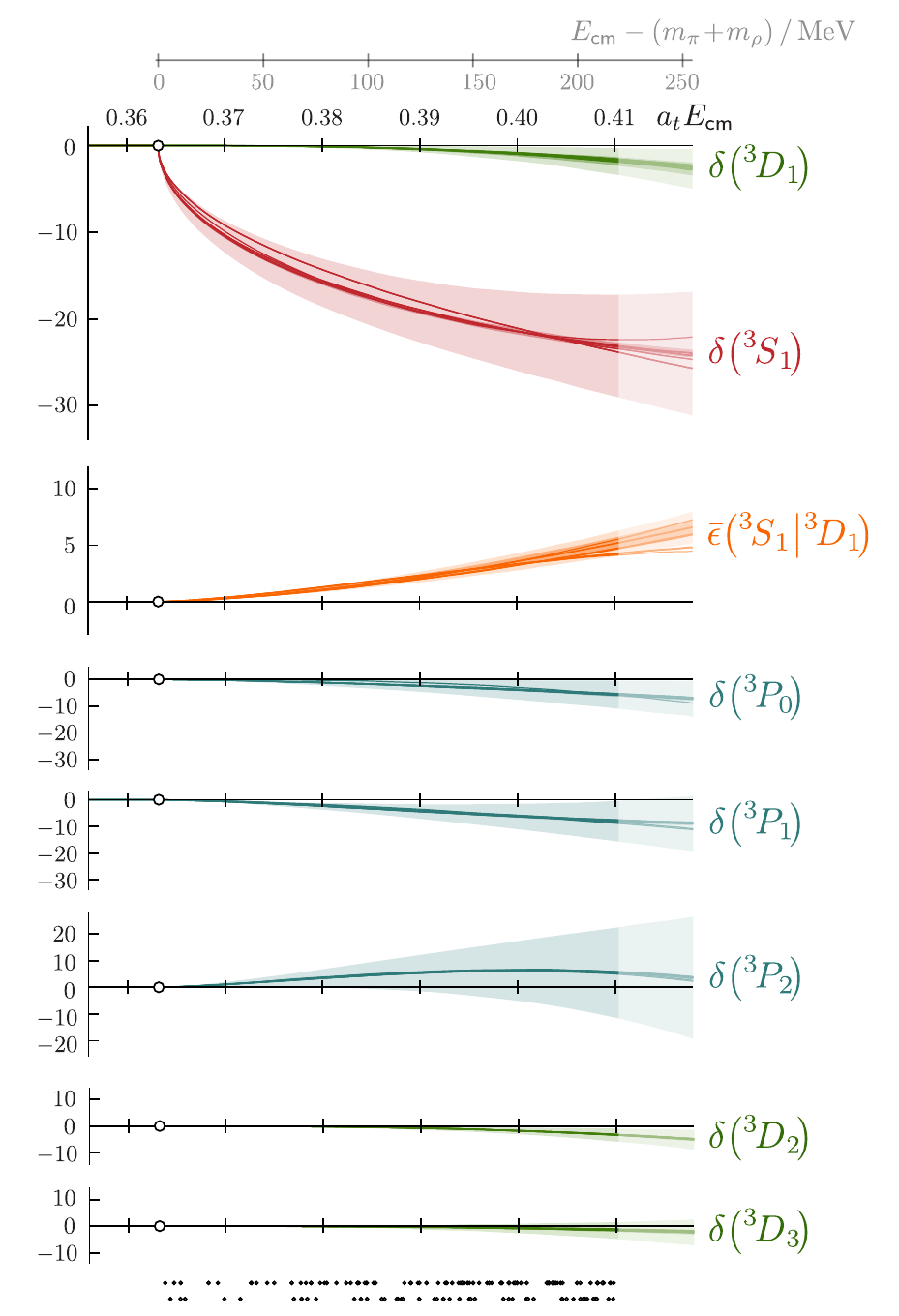}
	\caption{Phase shifts for partial-waves, $\threelJ$, and $\threeSone$ - $\threeDone$ mixing angle, $\bar{\epsilon}$, as described in the text. Each curve corresponds to a parameterisation in Table \ref{paramvar} with ${\chi^2/N_{\text{dof}} \leq 1.5}$. The darker inner band (typically thinner than the width of the curves) reflects the statistical uncertainty on the reference parameterisation in Table \ref{ALLparams} and the lighter outer bands correspond to the combined statistical and systematic uncertainties on this parameterisation. Faded regions highlight that no energy levels have been used to constrain the \mbox{phase-shifts} and mixing angle when $a_tE_\mathsf{cm}\geq 0.41$. The discrete energy levels used as constraints are shown as small dots at the bottom of the figure with the top and bottom rows for $L/a_s=24$ and $20$ respectively. An axis reflecting energy above threshold in physical units is displayed at the top of the figure.}
	\label{allplot}
\end{figure}

For all the parameterisations in Table~\ref{paramvar} with $\chi^2/N_{\text{dof}} \leq 1.5$, Figure~\ref{allplot} shows the two \mbox{phase-shifts} and mixing angle in the Stapp parameterisation, Eq.~(\ref{tinf}), for the dynamically-coupled $\threeSone$ and $\threeDone$ partial-waves, and the \mbox{phase-shifts} for the $\threePzero$, $\threePone$, $\threePtwo$, $\threeDtwo$ and $\threeDthree$ partial-waves. It can be seen that the scattering amplitudes are robust under varying the parameterisation with the \mbox{phase-shifts} consistent within statistical uncertainties. As expected, the systematic uncertainty, largely due to $\xi$ and hence discretisation effects, on each parameterisation dominates the uncertainty.

We conclude that $\rho\pi$ in isospin-2 is weakly repulsive in $\threeSone$. The other \mbox{phase-shifts} are consistent with zero within the systematic uncertainties, though there are hints of weak attraction in $\threePtwo$ and weak repulsion in $\threePzero$, $\threePone$ and $^3D_J$. The dynamical mixing between the $\threeSone$ and $\threeDone$-waves is small but significantly non-zero within the systematic uncertainties and across all parameterisations. In the following section we investigate in more detail how the spectra depend on the mixing angle.

\subsection{Constraints on the ${\threeSone}$ - ${\threeDone}$ mixing angle}\label{eps}
To demonstrate that the $\threeSone$ - $\threeDone$ mixing angle, $\bar{\epsilon}$, is being robustly constrained in the energy range considered, we investigate which energy levels are providing the most stringent constraints on it. If we neglect $\ell \geq 4$, the quantisation conditions, Eq.~(\ref{QCBSUB}), for irreps at rest admitting $\threeSone$, $\threeDone$-waves are independent of the sign of $\bar{\epsilon}$, whereas the quantisation conditions for irreps \mbox{in-flight} depend on the sign of $\bar{\epsilon}$. This means that for spatially periodic boundary conditions in a cubic box, ignoring contributions from $\ell \geq 4$, \mbox{in-flight} irreps must be considered in order to uniquely determine $\bar{\epsilon}$ from finite-volume spectra\footnote{If contributions of partial-waves with $\ell \geq 4$ are included for irreps overall at rest, then in general the finite-volume spectra are no longer independent of the sign of $\bar{\epsilon}$.}.

\begin{figure}
\centering
\includegraphics[scale=1]{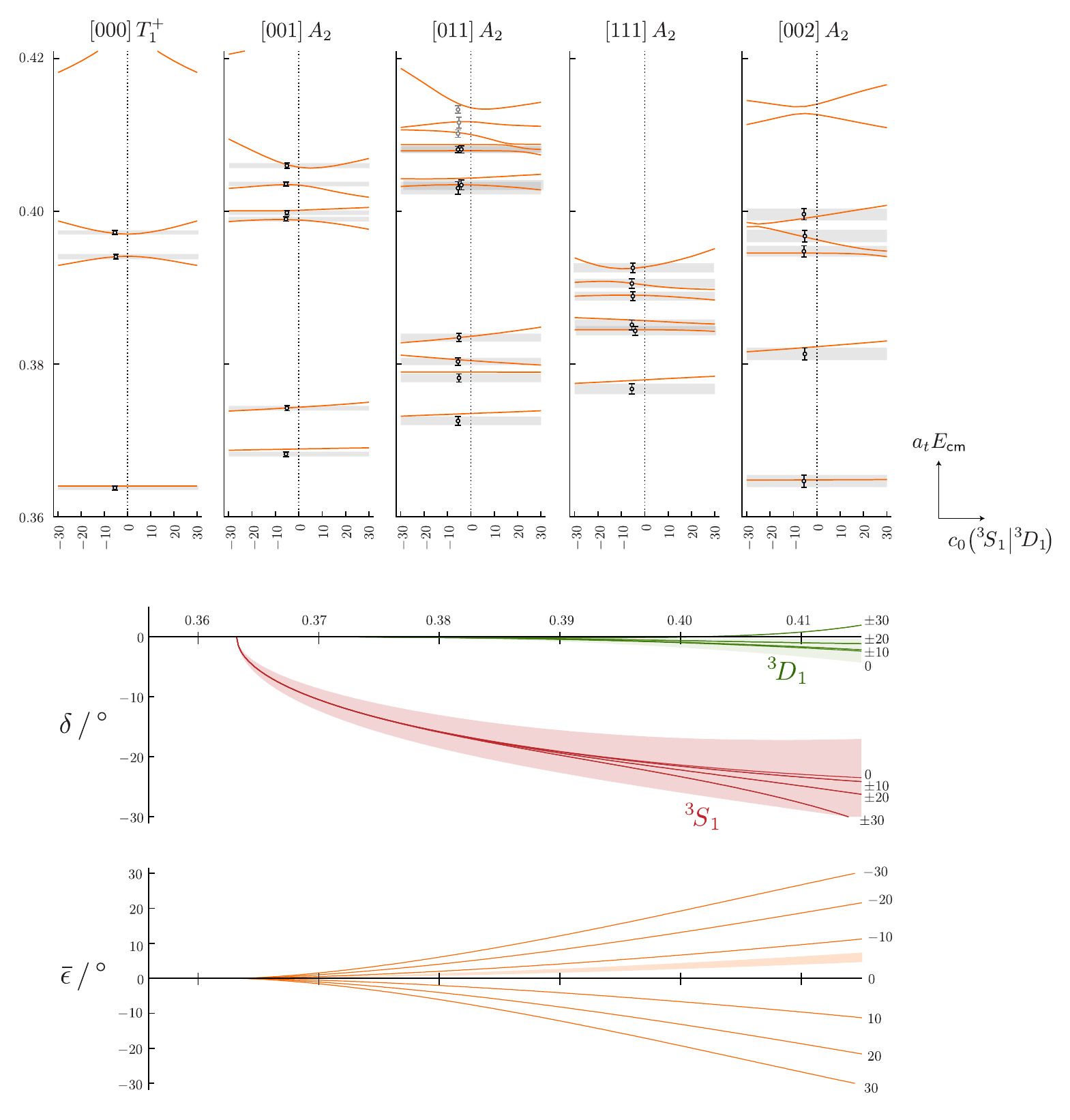}
\caption{Upper: Finite-volume spectra in the $[000]\,T_1^+$ and $\vec{P} \neq \vec{0}$ $A_2$ irreps on the $L/a_s=24$ lattice as a function of $c_0({\threeSone}|{\threeDone})$ as described in the text. Black and grey points are, as in Figure \ref{vm_p000}, energy levels extracted from correlation functions, plotted at $c_0({\threeSone}|{\threeDone})=-5.28$ the value in the reference amplitude parameterisation in Table~\ref{ALLparams}. Grey bands are to guide the eye and show the combined statistical and systematic uncertainties on the black points. Orange curves show the finite-volume spectra from the reference amplitude when $c_0({\threeSone}|{\threeDone})$ is varied with the other parameters fixed. Lower: $\delta_{\threeSone}(E_\mathsf{cm})$, $\delta_{\threeDone}(E_\mathsf{cm})$ and $\bar{\epsilon}(E_\mathsf{cm})$ for the reference amplitude with a selection of values for $c_0({\threeSone}|{\threeDone})$. The shaded bands shows the combined statistical and systematic uncertainties of the reference amplitudes, i.e.\ when $c_0({\threeSone}|{\threeDone})=-5.28$.}
\label{fv_vs_eta}
\end{figure}

Figure~\ref{fv_vs_eta} shows finite-volume spectra in the $[000]\,T_1^+$ irrep and the $\vec{P} \neq \vec{0}$ $A_2$ irreps as a function of the $\mathbf{K}$-matrix parameter $c_0({\threeSone}|{\threeDone})$ along with the corresponding \mbox{phase-shifts} $\delta_{\threeSone}$, $\delta_{\threeDone}$ and mixing angle $\bar{\epsilon}$.\footnote{The relations in Eq.~(\ref{tinf}) and Eq.~(\ref{ktinv}) can be manipulated to show that the sign of $c_0({\threeSone}|{\threeDone})$ is dependent on the sign of $\bar{\epsilon}$. The \mbox{phase-shifts} are independent of the sign of $c_0({\threeSone}|{\threeDone})$.}~The reference parameterisation in Table~\ref{ALLparams} has been used, varying $c_0({\threeSone}|{\threeDone})$ while keeping all other parameters fixed.  The symmetry of the finite-volume spectrum in $[000]\,T_1^+$ about $c_0({\threeSone}|{\threeDone})=0$ illustrates the expected sign independence at rest. For the $A_2$ irreps \mbox{in-flight}, the \mbox{finite-volume} spectra are clearly asymmetric about $c_0({\threeSone}|{\threeDone})=0$ and energy levels have a varying degree of dependence on $\bar{\epsilon}$. Furthermore, the \mbox{phase-shifts} vary only within their systematic uncertainties for ${-20\leq c_0({\threeSone}|{\threeDone})\leq 20}$, in stark contrast to $\bar{\epsilon}$. This suggests that the constraints placed on $c_0({\threeSone}|{\threeDone})$ by the finite-volume spectra are the most significant in determining $\bar{\epsilon}$ and Figure~\ref{fv_vs_eta} illustrates the numerous energy levels in the region $a_tE_\mathsf{cm}\leq0.41$ which provide these constraints, e.g.~the splitting between the $4^{th}$ and $5^{th}$ energy levels in the $[002]A_2$ irrep is strongly dependent on $c_0({\threeSone}|{\threeDone})$ in the small range we consider. Other irreps \mbox{in-flight} admitting the dynamically coupled $\threeSone$ and $\threeDone$ partial-waves provide additional constraints on $c_0({\threeSone}|{\threeDone})$ and subsequently $\bar{\epsilon}$. We conclude that these finite-volume calculations robustly determine the magnitude and sign of $\bar{\epsilon}$.

\section{Summary \label{Sec:Summary}}
In this paper we have reported on the first calculation of $\rho \pi$ scattering using lattice QCD, focusing on the isospin-2 channel. As expected for an exotic isospin, the hadron-hadron interactions are found to be relatively weak. The angular momentum barrier at low energy provides a natural hierarchy in $\ell$, and the coupling of $\ell$ with the intrinsic spin of the $\rho$ leads to a number of partial-waves for a given $J^P$. The possibility of `spin-orbit' forces in QCD allows amplitudes of common $\ell$, but distinct $J$, to differ. For each of $J^P=1^+, 2^- \ldots$ there are two dynamically-coupled partial-waves, and for $1^+$ we are able to determine the $\threeSone$ and $\threeDone$ amplitudes along with the coupling between them. We are also able to determine the scattering phase-shifts for all partial-waves of $\ell \leq 2$.

Our results followed from application of the formalism relating scattering amplitudes in an infinite volume to the discrete spectrum of QCD in a finite periodic volume defined by the lattice. We computed this spectrum in two spatial volumes in a version of QCD where the degenerate $u,d$ quarks are heavier than in experiment, such that they are degenerate with the strange quark and the theory has an exact SU(3) flavour symmetry. The resulting theory has octet pseudoscalar mesons (such as the $\pi$) of mass $\sim$ 700 MeV and stable octet vectors mesons (such as the $\rho$) of mass $\sim$ 1020 MeV. 

Spectra were obtained by variational analysis of matrices of two-point correlation functions computed using bases of operators resembling $\rho \pi$. The large number of partial-waves contributing, together with the weakness of the interactions, leads to spectra which feature many nearly-degenerate states. The use of bases of operators featuring all relevant `meson-meson' constructions in the energy region of interest leads to a robust determination, where the nearly degenerate states are resolved in the variational solution by virtue of their orthogonal overlap structures in the space of operators.

The spectra obtained in the two volumes, featuring 141 energy levels, were used to constrain the energy dependence of multiple partial-waves. Amplitudes were parameterised and the parameters adjusted so that the predicted finite-volume spectra matched the calculated spectra, as quantified by a correlated $\chi^2$. The dependence on the particular form of parameterisations used was explored and found to be rather modest. The largest single source of systematic uncertainty in the calculation was due to the difference in the lattice anisotropy for the $\pi$ and the various helicity components of the $\rho$. This is a relatively small discretisation effect, but its impact in this particular calculation is amplified by the weakness of the interactions -- this causes the finite-volume energy levels to be shifted relatively little from their non-interacting values.

The resulting scattering amplitudes presented in Figure~\ref{allplot} show a phase-shift in the $\threeSone$ channel which is clearly non-zero and repulsive. Phase-shifts for the other extracted partial waves are found to be compatible with zero within their systematic error. The mixing between $\threeSone$ and $\threeDone$ in $J^P=1^+$, as quantified by a mixing angle $\bar{\epsilon}$ in the Stapp parameterisation, is determined and found to be small but significantly \mbox{non-zero}. We are able to determine its sign by considering spectra where the $\rho\pi$ has overall non-zero momentum with respect to the lattice.

The low energy (near threshold) behaviour of the scattering amplitudes can be summarised in terms of the corresponding scattering lengths. Using the definition,\sloppy \  ${\lim_{k_\mathsf{cm} \to 0} k_\mathsf{cm}^{2\ell +1} \cot \big[\delta_{\threelJ} \big] = a({^3\ell_J}|{^3\ell_J})^{-1}}$, we find\footnote{We do not quote a $\threeDone$ scattering length because $t(\threeDone|\threeDone)\sim(k_\mathsf{cm})^4 \ll t(\threeSone|\threeDone)\sim(k_\mathsf{cm})^2$ at threshold and as such the contribution of $\bar{\epsilon}$ cannot be neglected, unlike in the $\threeSone$ case.},
\begin{alignat*}{2}
a(\threeSone|\!\threeSone)=& \,(-5.44 \pm 0.10 \pm 0.88) \cdot a_t &&\mspace{80mu}m_\pi\,a(\threeSone|\!\threeSone)= \, (-0.80 \pm 0.01 \pm 0.13) \\
a(\threePzero|\!\threePzero)=& \,(-132 \pm 14 \pm 104) \cdot a_t^3 &&\mspace{80mu}m_\pi^3\,a(\threePzero|\!\threePzero)= \, (-0.43 \pm 0.05 \pm 0.34) \\
a(\threePone|\!\threePone)=& \,(-303 \pm 12 \pm 114) \cdot a_t^3 &&\mspace{80mu}m_\pi^3\,a(\threePone|\!\threePone)= \, (-0.98 \pm 0.04 \pm 0.37)\\
a(\threePtwo|\!\threePtwo)=& \,(502 \pm 14 \pm 362) \cdot a_t^3 &&\mspace{80mu}m_\pi^3\,a(\threePtwo|\!\threePtwo)= \, ( 1.62 \pm 0.05 \pm 1.17) \\
a(\threeDtwo|\!\threeDtwo)=& \,(-8950 \pm 1050 \pm 5330) \cdot a_t^5 &&\mspace{80mu}m_\pi^5\,a(\threeDtwo|\!\threeDtwo)= \, (-0.63 \pm 0.07 \pm 0.38)\\
a(\threeDthree|\!\threeDthree)=& \,(-4320 \pm 1310 \pm 7270) \cdot a_t^5 &&\mspace{80mu}m_\pi^5\,a(\threeDthree|\!\threeDthree)= \, (-0.30 \pm 0.09 \pm 0.51) .
\end{alignat*}
The qualitative behaviour of the $^3P_J$-waves is the same as that found in Section~\ref{uncoupled-pwave} (where only irreps with a single non-negligible partial wave were considered) and each of the $^3P_J$ scattering lengths given above is consistent within errors with those found in Section~\ref{uncoupled-pwave}.

In conclusion, we have demonstrated how scattering amplitudes involving hadrons with non-zero spin can be computed using lattice QCD. Further applications of the approach presented here include the isospin-1 $\omega \pi$ system -- in the $J^P=1^+$ partial-wave this features a low-lying resonance, the $b_1$, which has been measured to have significant coupling\footnote{In cases of channels featuring stronger interactions, and in particular those including resonances, we expect the relative uncertainty due to discretisation effects felt through the anisotropy to be much reduced.} to both $\threeSone$ and $\threeDone$ channels~\cite{Nozar:2002br}. Furthermore, contemporary experiments in the charmonium sector appear to show resonant behaviour in the exotic-flavour $J/\psi \, \pi$ channel; first attempts to determine lattice QCD spectra here have appeared~\cite{Cheung:2017tnt,Prelovsek:2014swa}, but as yet there has been no determination of the scattering amplitudes.

\acknowledgments
{
	We thank our colleagues within the Hadron Spectrum Collaboration, with particular thanks to R.~Brice\~no for useful discussions.
	AJW is supported by the U.K. Science and Technology Facilities Council
	(STFC).  AJW and CET acknowledge support from STFC [grant number ST/P000681/1]. 
	JJD acknowledges support from the U.S. Department of Energy Early Career award contract de-sc0006765.
	JJD and RGE acknowledge support from the U.S. Department of Energy contract DE-AC05-06OR23177, under which Jefferson Science Associates, LLC, manages and operates Jefferson Lab. 
	DJW acknowledges support from a Royal Society--Science Foundation Ireland University Research Fellowship Award UF160419 and from the European Union’s Horizon 2020 research and innovation programme under grant Agreement No. 749850--XXQCD.
	
	The contractions were performed on clusters at Jefferson Lab under the USQCD Collaboration and the Scientific Discovery through Advanced Computing (SciDAC) program.
	The software codes {\tt Chroma}~\cite{Edwards:2004sx} and {\tt QUDA}~\cite{Clark:2009wm,Babich:2010mu} were used to compute the propagators required for this project.
	This research used resources of the National Energy Research Scientific Computing Center (NERSC), a DOE Office of Science User Facility supported by the Office of Science of the U.S. Department of Energy under Contract No. DEAC02-05CH11231. The authors acknowledge the Texas Advanced Computing Center (TACC) at The University of Texas at Austin for providing HPC resources that have contributed to the research results reported within this paper.
	Gauge configurations were generated using resources awarded from the U.S. Department of Energy INCITE program at the Oak Ridge Leadership Computing Facility, the NERSC, the NSF Teragrid at the TACC and the Pittsburgh Supercomputer Center, as well as at Jefferson Lab.
}

\appendix

\section{Subduction of vector-pseudoscalar partial-waves for $\vec{P}\neq \vec{0}$ \label{App:Subduction_In_Flight}}
Tables~\ref{tab001},~\ref{tab011} and~\ref{tab111} present the subduction patterns for vector-pseudoscalar partial-waves with $\ell \le 3$ for momenta of type $[00n], [0nn]$ and $[nnn]$ respectively for integer $n$.

\begin{table}[h!] 
{\renewcommand{\arraystretch}{1.2}
\begin{tabular}{ c |l l l l l} 
$[00n]\, \Lambda$ & \multicolumn{1}{c}{$A_1$} &  \multicolumn{1}{c}{$A_2$} &  \multicolumn{1}{c}{$E$} &  \multicolumn{1}{c}{$B_1$} &  \multicolumn{1}{c}{$B_2$} \\ 
\hline 
&&&&&\\[-2.5ex]
\multirow{12}{*}{$J^P( \threelJ)$} & & $0^- \, \left( \threePzero \right)$ &&&\\[0.5ex]
&& $1^+ \left( \begin{matrix} \threeSone \\  \threeDone \end{matrix}   \right)$ & $1^+ \left( \begin{matrix} \threeSone \\  \threeDone \end{matrix}   \right)$ &&\\[0.5ex]
& $1^- \, \left( \threePone \right)$ && $1^- \, \left( \threePone \right)$ &&\\[0.5ex]
& $2^+ \, \left( \threeDtwo \right)$ & & $2^+ \, \left( \threeDtwo \right)$ & $2^+ \, \left( \threeDtwo \right)$ & $2^+ \, \left( \threeDtwo \right)$ \\[0.5ex]
&& $2^- \left( \begin{matrix} \threePtwo \\  \threeFtwo \end{matrix}   \right)$ & $2^- \left( \begin{matrix} \threePtwo \\  \threeFtwo \end{matrix}   \right)$ & $2^- \left( \begin{matrix} \threePtwo \\  \threeFtwo \end{matrix}   \right)$ & $2^- \left( \begin{matrix} \threePtwo \\  \threeFtwo \end{matrix}   \right)$  \\[0.5ex]
&& $3^+ \, \left( \begin{matrix} \threeDthree \\  {\color{gray}\threeGthreeit} \end{matrix}   \right)$ & $3^+ \, \left( \begin{matrix} \threeDthree \\  {\color{gray}\threeGthreeit} \end{matrix}   \right)_{[2]}$ & $3^+ \, \left( \begin{matrix} \threeDthree \\  {\color{gray}\threeGthreeit} \end{matrix}   \right)$ & $3^+ \, \left( \begin{matrix} \threeDthree \\  {\color{gray}\threeGthreeit} \end{matrix}   \right)$ \\[0.5ex]
& $3^- \, \left( \threeFthree \right)$ && $3^- \, \left( \threeFthree \right)_{[2]}$ & $3^- \, \left( \threeFthree \right)$ & $3^- \, \left( \threeFthree \right)$ \\[0.5ex]
& $4^- \, \left( \begin{matrix} \threeFfour \\  {\color{gray}\threeHfourit} \end{matrix}   \right)$ & $4^- \, \left( \begin{matrix} \threeFfour \\  {\color{gray}\threeHfourit} \end{matrix}   \right)_{[2]}$ & $4^- \, \left( \begin{matrix} \threeFfour \\  {\color{gray}\threeHfourit} \end{matrix}   \right)_{[2]}$ & $4^- \, \left( \begin{matrix} \threeFfour \\  {\color{gray}\threeHfourit} \end{matrix}   \right)$ & $4^- \, \left( \begin{matrix} \threeFfour \\  {\color{gray}\threeHfourit} \end{matrix}   \right)$ \\[0.5ex]
\hline
\end{tabular}
}
\caption{Partial-wave $J^P(\threelJ)$ subductions for $\ell \le 3$ at $\vec{P}=[00n]$ into irreps $\Lambda$ of the little-group $\text{Dic}_4$. A subscript $[N]$ indicates that this $J^P$ has $N$ embeddings in the irrep $\Lambda$. Partial-waves with $\ell > 3$ that couple dynamically to partial-waves with $\ell \leq 3$ are shown in grey italic. This table is derived using the results presented in Refs.~\cite{Moore:2006ng} and \cite{Thomas:2011rh}.} \label{tab001}
\end{table}
\begin{table}[h!] 
{\renewcommand{\arraystretch}{1.2}
\begin{tabular}{ c |l l l l } 
$[0nn]\, \Lambda$ & \multicolumn{1}{c}{$A_1$} &  \multicolumn{1}{c}{$A_2$} & 
 \multicolumn{1}{c}{$B_1$} &  \multicolumn{1}{c}{$B_2$} \\ 
\hline 
&&&&\\[-2.5ex]
\multirow{12}{*}{$J^P( \threelJ)$} & & $0^- \, \left( \threePzero \right)$ &&\\[0.5ex]
&& $1^+ \left( \begin{matrix} \threeSone \\  \threeDone \end{matrix}   \right)$ & $1^+ \left( \begin{matrix} \threeSone \\  \threeDone \end{matrix}   \right)$& $1^+ \left( \begin{matrix} \threeSone \\  \threeDone \end{matrix}   \right)$\\[0.5ex]
& $1^- \, \left( \threePone \right)$ && $1^- \, \left( \threePone \right)$ & $1^- \, \left( \threePone \right)$ \\[0.5ex]
& $2^+ \, \left( \threeDtwo \right)_{[2]}$  & $2^+ \, \left( \threeDtwo \right)$ & $2^+ \, \left( \threeDtwo \right)$ & $2^+ \, \left( \threeDtwo \right)$ \\[0.5ex]
& $2^- \left( \begin{matrix} \threePtwo \\  \threeFtwo \end{matrix}   \right)$ & $2^- \left( \begin{matrix} \threePtwo \\  \threeFtwo \end{matrix}   \right)_{[2]}$ & $2^- \left( \begin{matrix} \threePtwo \\  \threeFtwo \end{matrix}   \right)$ & $2^- \left( \begin{matrix} \threePtwo \\  \threeFtwo \end{matrix}   \right)$  \\[0.5ex]
& $3^+ \, \left( \begin{matrix} \threeDthree \\  {\color{gray}\threeGthreeit} \end{matrix}   \right)$ & $3^+ \, \left( \begin{matrix} \threeDthree \\  {\color{gray}\threeGthreeit} \end{matrix}   \right)_{[2]}$ & $3^+ \, \left( \begin{matrix} \threeDthree \\  {\color{gray}\threeGthreeit} \end{matrix}   \right)_{[2]}$ & $3^+ \, \left( \begin{matrix} \threeDthree \\  {\color{gray}\threeGthreeit} \end{matrix}   \right)_{[2]}$ \\[0.5ex]
& $3^- \, \left( \threeFthree \right)_{[2]}$ & $3^- \, \left( \threeFthree \right)$ & $3^- \, \left( \threeFthree \right)_{[2]}$ & $3^- \, \left( \threeFthree \right)_{[2]}$ \\[0.5ex]
& $4^- \, \left( \begin{matrix} \threeFfour \\  {\color{gray}\threeHfourit} \end{matrix}   \right)_{[2]}$ & $4^- \, \left( \begin{matrix} \threeFfour \\  {\color{gray}\threeHfourit} \end{matrix}   \right)_{[3]}$ & $4^- \, \left( \begin{matrix} \threeFfour \\  {\color{gray}\threeHfourit} \end{matrix}   \right)_{[2]}$ & $4^- \, \left( \begin{matrix} \threeFfour \\  {\color{gray}\threeHfourit} \end{matrix}   \right)_{[2]}$  \\[0.5ex]
\hline
\end{tabular}
}
\caption{As Table~\ref{tab001}, but for $\vec{P}=[0nn]$ with little-group $\mathrm{Dic}_2$. } \label{tab011}
\end{table}
\begin{table}[h!] 
{\renewcommand{\arraystretch}{1.2}
\begin{tabular}{ c |l  l l } 
$[nnn]\, \Lambda$ & \multicolumn{1}{c}{$A_1$} &  \multicolumn{1}{c}{$A_2$} & 
 \multicolumn{1}{c}{$E$}  \\ 
\hline 
&&&\\[-2.5ex]
\multirow{12}{*}{$J^P( \threelJ)$} & & $0^- \, \left( \threePzero \right)$ &\\[0.5ex]
&& $1^+ \left( \begin{matrix} \threeSone \\  \threeDone \end{matrix}   \right)$ & $1^+ \left( \begin{matrix} \threeSone \\  \threeDone \end{matrix}   \right)$  \\[0.5ex]
& $1^- \, \left( \threePone \right)$ && $1^- \, \left( \threePone \right)$  \\[0.5ex]
& $2^+ \, \left( \threeDtwo \right)$  && $2^+ \, \left( \threeDtwo \right)_{[2]}$ \\[0.5ex]
&& $2^- \left( \begin{matrix} \threePtwo \\  \threeFtwo \end{matrix}   \right)$ & $2^- \left( \begin{matrix} \threePtwo \\  \threeFtwo \end{matrix}   \right)_{[2]}$   \\[0.5ex]
& $3^+ \, \left( \begin{matrix} \threeDthree \\  {\color{gray}\threeGthreeit} \end{matrix}   \right)$ & $3^+ \, \left( \begin{matrix} \threeDthree \\  {\color{gray}\threeGthreeit} \end{matrix}   \right)_{[2]}$ & $3^+ \, \left( \begin{matrix} \threeDthree \\  {\color{gray}\threeGthreeit} \end{matrix}   \right)_{[2]}$ \\[0.5ex]
& $3^- \, \left( \threeFthree \right)_{[2]}$ & $3^- \, \left( \threeFthree \right)$ & $3^- \, \left( \threeFthree \right)_{[2]}$ \\[0.5ex]
& $4^- \, \left( \begin{matrix} \threeFfour \\  {\color{gray}\threeHfourit} \end{matrix}   \right)$ & $4^- \, \left( \begin{matrix} \threeFfour \\  {\color{gray}\threeHfourit} \end{matrix}   \right)_{[2]}$ & $4^- \, \left( \begin{matrix} \threeFfour \\  {\color{gray}\threeHfourit} \end{matrix}   \right)_{[3]}$  \\[0.5ex]
\hline
\end{tabular}
}
\caption{As Table~\ref{tab001}, but for $\vec{P}=[nnn]$ with little-group $\mathrm{Dic}_3$. } \label{tab111}
\end{table}

\section{Finite-volume spectra \label{App:Finite_Volume_Spectra}}
\begin{figure}[tb]     
\captionsetup[subfigure]{justification=centering}
	\centering
	\includegraphics[scale=0.8875]{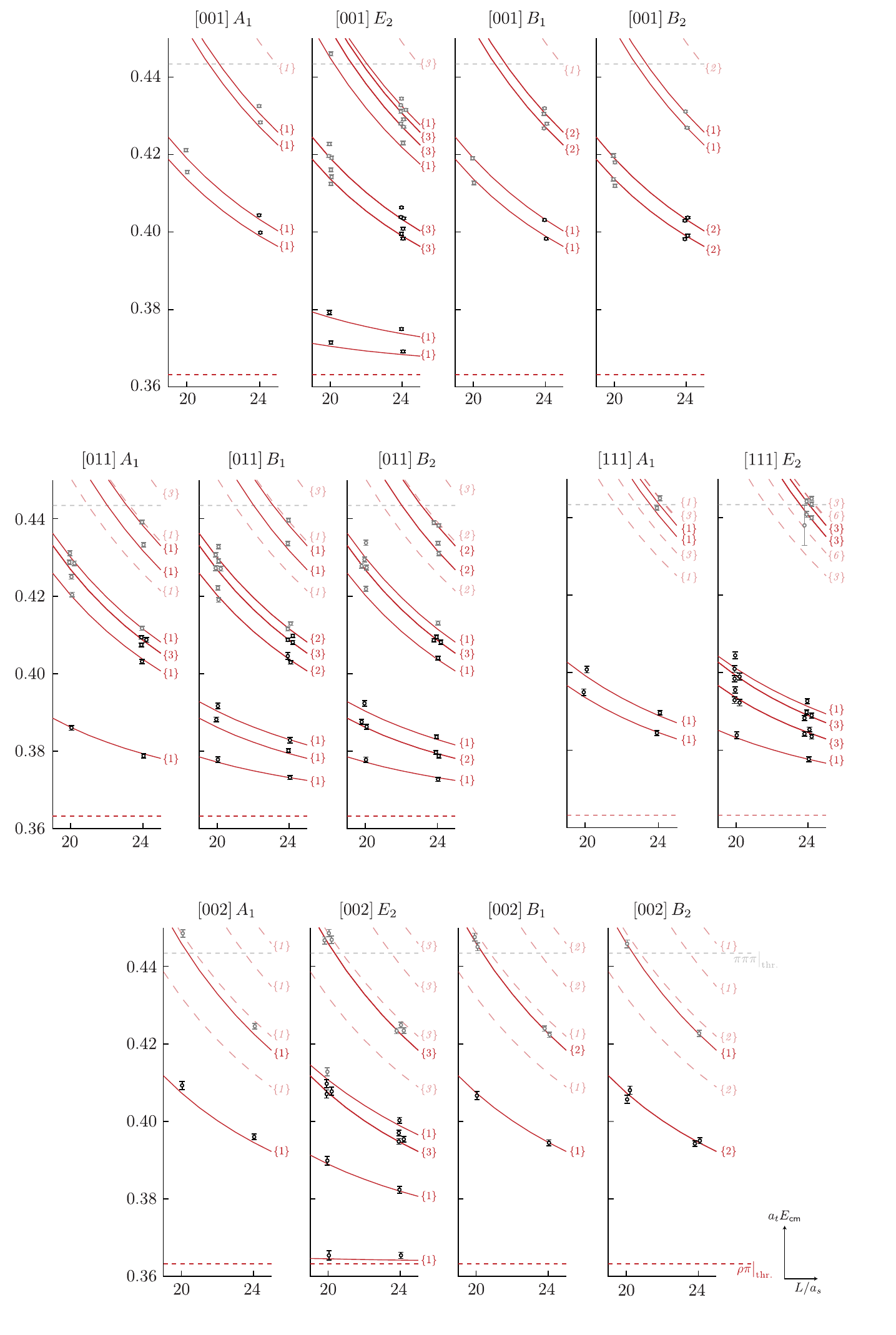}
	\caption{As Figure \ref{vm_DA2} but for all other irreps with $|\vec{P}|^2\leq 4$.}
	\label{vm_app}
\end{figure}

We provide here the finite-volume spectra plots for irreps at non-zero momenta, not shown in Figures~\ref{vm_p000} and~\ref{vm_DA2}, in Figure~\ref{vm_app}.  We also show the operator basis in Tables~\ref{optabB1} -~\ref{optabB4} for all irreps considered in Figures~\ref{vm_p000},~\ref{vm_DA2} and~\ref{vm_app} that were not shown in Table~\ref{i2_op_tab}.

\begin{table}[tb]
	\small
	\begin{tabular}{r : r : r : r : r}
		\multicolumn{1}{c:}{$[000]A_2^+$} 
		&  \multicolumn{1}{c:}{$[000]E^+$ } 
		&  \multicolumn{1}{c:}{$[000]T_2^+$} 
		&  \multicolumn{1}{c:}{$[000]A_1^-$} 
		&  \multicolumn{1}{c}{$[000]T_1^-$} \\[0.5ex]
		\cmidrule(lr){1-5}
		$\rho_{[011]}\pi_{[0\text{-}1\text{-}1]}$ & $\rho_{[011]}\pi_{[0\text{-}1\text{-}1]}$ & $\rho_{[001]}\pi_{[00\text{-}1]}$ & $\rho_{[001]}\pi_{[00\text{-}1]}$ & $\rho_{[001]}\pi_{[00\text{-}1]}$ \\[0.5ex] 
		$\rho_{[111]}\pi_{[\text{-}1\text{-}1\text{-}1]}$ & $\rho_{[111]}\pi_{[\text{-}1\text{-}1\text{-}1]}$ & $\{2\}\;\rho_{[011]}\pi_{[0\text{-}1\text{-}1]}$  & $\rho_{[011]}\pi_{[0\text{-}1\text{-}1]}$  & $\{2\}\;\rho_{[011]}\pi_{[0\text{-}1\text{-}1]}$  \\[0.5ex]
		 & & $\rho_{[111]}\pi_{[\text{-}1\text{-}1\text{-}1]}$  & $\rho_{[111]}\pi_{[\text{-}1\text{-}1\text{-}1]}$  & $\rho_{[111]}\pi_{[\text{-}1\text{-}1\text{-}1]}$\\[0.5ex]
	\cmidrule(lr){1-5}
		$2\,\text{ops.}$ & $2\,\text{ops.}$ & $4\,\text{ops.}$ & $3\,\text{ops.}$ & $4\,\text{ops.}$ \\
	\end{tabular}
	\caption{As Table \ref{i2_op_tab} but for irreps $A_2^+$, $E^+$, $T_2^+$, $A_1^-$ and $T_1^-$ at $\vec{P}=[000]$.}
	\label{optabB1}
\end{table}
\begin{table}[tb]
	\small
	\begin{tabular}{r : r : r : r : r : r}
		\multicolumn{1}{c:}{$[000]E^-$} 
		&  \multicolumn{1}{c:}{$[000]T_2^-$ } 
		&
		&  \multicolumn{1}{c:}{$[001]A_1$} 
		&  \multicolumn{1}{c:}{$[001]B_1$} 
		&  \multicolumn{1}{c}{$[001]B_2$} \\[0.5ex]
		\cmidrule(lr){1-2}\cmidrule(lr){4-6}
		$\rho_{[001]}\pi_{[00\text{-}1]}$ & $\rho_{[001]}\pi_{[00\text{-}1]}$ 
		& & $\rho_{[011]}\pi_{[0\text{-}10]}$ 
		& $\rho_{[011]}\pi_{[0\text{-}10]}$
		& $\{2\}\;\rho_{[011]}\pi_{[0\text{-}10]}$
		 \\[0.5ex] 
		$\{2\}\;\rho_{[011]}\pi_{[0\text{-}1\text{-}1]}$ & $\{2\}\;\rho_{[011]}\pi_{[0\text{-}1\text{-}1]}$ & & $\rho_{[010]}\pi_{[0\text{-}11]}$ & $\rho_{[010]}\pi_{[0\text{-}11]}$  & $\{2\}\;\rho_{[010]}\pi_{[0\text{-}11]}$\\[0.5ex]
		$\rho_{[111]}\pi_{[\text{-}1\text{-}1\text{-}1]}$ & $\{2\}\;\rho_{[111]}\pi_{[\text{-}1\text{-}1\text{-}1]}$ & &
		$\rho_{[111]}\pi_{[\text{-}1\text{-}10]}$ &  
		$\{2\}\;\rho_{[111]}\pi_{[\text{-}1\text{-}10]}$ & $\rho_{[111]}\pi_{[\text{-}1\text{-}10]}$\\[0.5ex]
	    & & & 
		$\rho_{[110]}\pi_{[\text{-}1\text{-}11]}$ &  
		$\{2\}\;\rho_{[110]}\pi_{[\text{-}1\text{-}11]}$ & $\rho_{[110]}\pi_{[\text{-}1\text{-}11]}$\\[0.5ex]
		& & & 
		{\color{gray}$\mathit{\rho_{[012]}\pi_{[0\text{-}1\text{-}1]}}$} &  
		{\color{gray}$\mathit{\rho_{[012]}\pi_{[0\text{-}1\text{-}1]}}$}& {\color{gray}$\mathit{\{2\}\;\rho_{[012]}\pi_{[0\text{-}1\text{-}1]}}$}\\[0.5ex]
		\cmidrule(lr){1-2}\cmidrule(lr){4-6}
		$4\,\text{ops.}$ & $5\,\text{ops.}$ & & $4\,\text{ops.}$ & $6\,\text{ops.}$ & $6\,\text{ops.}$ \\
	\end{tabular}
	\caption{As Table \ref{i2_op_tab} but for irreps $E^-$ and $T_2^-$ at $\vec{P}=[000]$ and $A_1$, $B_1$ and $B_2$ at $\vec{P}=[001]$.}
	\label{optabB2}
\end{table}
\begin{table}[tb]
	\small
	\begin{tabular}{r : r : r : r : r : r : r}
		\multicolumn{1}{c:}{$[001]E_2$} 
		&
		&  \multicolumn{1}{c:}{$[011]A_1$ } 
		&  \multicolumn{1}{c:}{$[011]B_1$} 
		&  \multicolumn{1}{c:}{$[011]B_2$} 
		&
		&  \multicolumn{1}{c}{$[111]A_1$} \\[0.5ex]
		\cmidrule(lr){1-1}\cmidrule(lr){3-5}\cmidrule(lr){7-7}
		$\rho_{[001]}\pi_{[000]}$ & & 
		$\rho_{[001]}\pi_{[010]}$ 
		& $\rho_{[011]}\pi_{[000]}$ 
		& $\rho_{[011]}\pi_{[000]}$ &
		& $\rho_{[011]}\pi_{[100]}$
		\\[0.5ex] 
		$\rho_{[000]}\pi_{[001]}$ & &  
		$\rho_{[111]}\pi_{[\text{-}100]}$ & 
		$\rho_{[001]}\pi_{[010]}$ & 
		$\{2\}\;\rho_{[001]}\pi_{[010]}$  & &
		$\rho_{[001]}\pi_{[110]}$\\[0.5ex]
		$\{3\}\;\rho_{[011]}\pi_{[0\text{-}10]}$ & &  $\{3\}\;\rho_{[110]}\pi_{[\text{-}101]}$ & 
		$\rho_{[000]}\pi_{[011]}$ & 
		$\rho_{[000]}\pi_{[011]}$ & &
		{\color{gray}$\mathit{\rho_{[112]}\pi_{[00\text{-}1]}}$}\\[0.5ex]
		$\{3\}\;\rho_{[010]}\pi_{[0\text{-}11]}$ & &
		$\rho_{[100]}\pi_{[\text{-}111]}$ & 
		$\{2\}\;\rho_{[111]}\pi_{[\text{-}100]}$ & 
		$\rho_{[111]}\pi_{[\text{-}100]}$  & &
		{\color{gray}$\mathit{\{3\}\;\rho_{[012]}\pi_{[10\text{-}1]}}$}\\[0.5ex]
		$\rho_{[002]}\pi_{[00\text{-}1]}$ & & 
		{\color{gray}$\mathit{\rho_{[012]}\pi_{[00\text{-}1]}}$} & 
        $\{3\}\;\rho_{[110]}\pi_{[\text{-}101]}$ & 
        $\{3\}\;\rho_{[110]}\pi_{[\text{-}101]}$ & &
        $\rho_{[002]}\pi_{[11\text{-}1]}$\\[0.5ex]
		$\{3\}\;\rho_{[111]}\pi_{[\text{-}1\text{-}10]}$ & &
		$\rho_{[002]}\pi_{[01\text{-}1]}$ & 
		$\{2\}\;\rho_{[100]}\pi_{[\text{-}111]}$  & 
		$\rho_{[100]}\pi_{[\text{-}111]}$ & &
		$\rho_{[11\text{-}1]}\pi_{[002]}$\\[0.5ex]
		$\{3\}\;\rho_{[110]}\pi_{[\text{-}1\text{-}11]}$ & &
		$\rho_{[01\text{-}1]}\pi_{[002]}$ & 
		{\color{gray}$\mathit{\rho_{[012]}\pi_{[00\text{-}1]}}$} & 
		{\color{gray}$\mathit{\{2\}\;\rho_{[012]}\pi_{[00\text{-}1]}}$} & & {\color{gray}$\mathit{\{3\}\;\rho_{[10\text{-}1]}\pi_{[012]}}$}\\[0.5ex]
		$\rho_{[00\text{-}1]}\pi_{[002]}$ & &
		{\color{gray}$\mathit{\rho_{[00\text{-}1]}\pi_{[012]}}$} & 
        $\rho_{[002]}\pi_{[01\text{-}1]}$  & 
		$\{2\}\;\rho_{[002]}\pi_{[01\text{-}1]}$  & &
		{\color{gray}$\mathit{\rho_{[00\text{-}1]}\pi_{[112]}}$}\\[0.5ex]
		{\color{gray}$\mathit{\{3\}\;\rho_{[012]}\pi_{[0\text{-}1\text{-}1]}}$}&& 
		{\color{gray}$\mathit{\{3\}\;\rho_{[\text{-}10\text{-}1]}\pi_{[112]}}$} & 
		$\rho_{[01\text{-}1]}\pi_{[002]}$  & 
		$\{2\}\;\rho_{[01\text{-}1]}\pi_{[002]}$ & &
		\\[0.5ex]
		& & 
		& 
		{\color{gray}$\mathit{\rho_{[00\text{-}1]}\pi_{[012]}}$}  & 
		{\color{gray}$\mathit{\{2\}\;\rho_{[00\text{-}1]}\pi_{[012]}}$} & &
		\\[0.5ex]
		& & 
		& 
		{\color{gray}$\mathit{\{3\}\;\rho_{[112]}\pi_{[\text{-}10\text{-}1]}}$}  & 
		{\color{gray}$\mathit{\{3\}\;\rho_{[112]}\pi_{[\text{-}10\text{-}1]}}$}& &
		\\[0.5ex]
		\cmidrule(lr){1-1}\cmidrule(lr){3-5}\cmidrule(lr){7-7}
		$16\,\text{ops.}$ & & $8\,\text{ops.}$ & $12\,\text{ops.}$ & $13\,\text{ops.}$ & & $4\,\text{ops.}$ \\
	\end{tabular}
	\caption{As Table \ref{i2_op_tab} but for irreps $E_2$ at $\vec{P}=[001]$; $A_1$, $B_1$ and $B_2$ at $\vec{P}=[011]$ and $A_1$ at $\vec{P}=[111]$.}
	\label{optabB3}
\end{table}
\begin{table}[tb]
	\small
	\begin{tabular}{r : r : r : r : r  : r}
		\multicolumn{1}{c:}{$[111]E_2$} 
		&
		&  \multicolumn{1}{c:}{$[002]A_1$ } 
		&  \multicolumn{1}{c:}{$[002]B_1$} 
		&  \multicolumn{1}{c:}{$[002]B_2$} 
		&  \multicolumn{1}{c}{$[002]E_2$} \\[0.5ex]
		\cmidrule(lr){1-1}\cmidrule(lr){3-6}
		$\rho_{[111]}\pi_{[000]}$ & & 
		$\rho_{[011]}\pi_{[0\text{-}11]}$ 
		& $\rho_{[011]}\pi_{[0\text{-}11]}$
		& $\{2\}\;\rho_{[011]}\pi_{[0\text{-}11]}$
		& $\rho_{[001]}\pi_{[001]}$
		\\[0.5ex]
		$\{3\}\;\rho_{[011]}\pi_{[100]}$ & & 
		{\color{gray}$\mathit{\rho_{[012]}\pi_{[0\text{-}10]}}$}
		& {\color{gray}$\mathit{\rho_{[012]}\pi_{[0\text{-}10]}}$}
		& {\color{gray}$\mathit{\{2\}\;\rho_{[012]}\pi_{[0\text{-}10]}}$}
		& $\rho_{[002]}\pi_{[000]}$
		\\[0.5ex]  
		$\{3\}\;\rho_{[100]}\pi_{[011]}$ & & 
		$\rho_{[111]}\pi_{[\text{-}1\text{-}11]}$ 
		& $\{2\}\;\rho_{[111]}\pi_{[\text{-}1\text{-}11]}$ 
		& $\rho_{[111]}\pi_{[\text{-}1\text{-}11]}$ 
		& $\{3\}\;\rho_{[011]}\pi_{[0\text{-}11]}$
		\\[0.5ex]
		$\rho_{[000]}\pi_{[111]}$ & & 
		{\color{gray}$\mathit{\rho_{[0\text{-}10]}\pi_{[012]}}$}
		& {\color{gray}$\mathit{\rho_{[0\text{-}10]}\pi_{[012]}}$}
		& {\color{gray}$\mathit{\{2\}\;\rho_{[0\text{-}10]}\pi_{[012]}}$}
		& $\rho_{[000]}\pi_{[002]}$
		\\[0.5ex]
		{\color{gray}$\mathit{\{3\}\;\rho_{[112]}\pi_{[00\text{-}1]}}$} & & 
		{\color{gray}$\mathit{\rho_{[112]}\pi_{[\text{-}1\text{-}10]}}$}
		& {\color{gray}$\mathit{\{2\}\;\rho_{[112]}\pi_{[\text{-}1\text{-}10]}}$}
		& {\color{gray}$\mathit{\rho_{[112]}\pi_{[\text{-}1\text{-}10]}}$}
		& {\color{gray}$\mathit{\{3\}\;\rho_{[012]}\pi_{[0\text{-}10]}}$}
		\\[0.5ex]
		{\color{gray}$\mathit{\{6\}\;\rho_{[012]}\pi_{[10\text{-}1]}}$} & & 
		{\color{gray}$\mathit{\rho_{[110]}\pi_{[\text{-}1\text{-}12]}}$}
		& 	{\color{gray}$\mathit{\{2\}\;\rho_{[110]}\pi_{[\text{-}1\text{-}12]}}$}
		& 	{\color{gray}$\mathit{\rho_{[110]}\pi_{[\text{-}1\text{-}12]}}$}
		& $\{3\}\;\rho_{[111]}\pi_{[\text{-}1\text{-}11]}$ 
		\\[0.5ex]
		$\{3\}\;\rho_{[002]}\pi_{[11\text{-}1]}$ & & 
		& 
		& 
		& {\color{gray}$\mathit{\{3\}\;\rho_{[0\text{-}10]}\pi_{[012]}}$}
		\\[0.5ex]
		$\{3\}\;\rho_{[11\text{-}1]}\pi_{[002]}$ & & 
		& 
		& 
		& {\color{gray}$\mathit{\{3\}\;\rho_{[112]}\pi_{[\text{-}1\text{-}10]}}$}
		\\[0.5ex]
	    {\color{gray}$\mathit{\{6\}\;\rho_{[01\text{-}1]}\pi_{[102]}}$}  & & 
		& 
		& 
		& {\color{gray}$\mathit{\{3\}\;\rho_{[110]}\pi_{[\text{-}1\text{-}12]}}$}
		\\[0.5ex]
		{\color{gray}$\mathit{\{3\}\;\rho_{[00\text{-}1]}\pi_{[112]}}$}  & & 
		& 
		& 
		&
		\\[0.5ex]
    	\cmidrule(lr){1-1}\cmidrule(lr){3-6}
		$14\,\text{ops.}$ & & $2\,\text{ops.}$ & $3\,\text{ops.}$ & $3\,\text{ops.}$ & $9\,\text{ops.}$ \\
	\end{tabular}
	\caption{As Table \ref{i2_op_tab} but for irreps $E_2$ at $\vec{P}=[111]$ and $A_1$, $B_1$, $B_2$ and $E_2$ at $\vec{P}=[002]$.}
	\label{optabB4}
\end{table}

\section{Details of the quantisation condition \label{App:Details_Of_Luescher}}
The quantisation condition relating infinite-volume scattering amplitudes to the finite-volume spectrum in a periodic $L\times L \times L$ box can be constructed from Equation~(22) of Ref.~\cite{Briceno:2014oea}. In the case of a single channel of vector-pseudoscalar scattering it can be written
\begin{equation}
\det_{\ell J m}\big[ \mathbf{1} + i \,\rho \,\mathbf{t}\,  \big( \mathbf{1} + i \overline{\boldsymbol{\mathcal{M}}} \big) \big] = 0,
\end{equation}
where the transcription of notation, $\big(\mathcal{M}\big)^{\text{\cite{Briceno:2014oea}}} = 16 \pi  \mathbf{t}$ and ${\big(\delta G^V \big)^{\text{\cite{Briceno:2014oea}}} = \tfrac{i}{16\pi} \rho \big( \mathbf{1} + i \overline{\boldsymbol{\mathcal{M}}} \big)}$
refers to the quantities defined in Ref.~\cite{Briceno:2014oea}. The resulting matrix of finite-volume functions is 
\begin{align}
\overline{\mathcal{M}}_{\ell J m,\,  \ell' J' m'} = 
\sum_{m_\ell, m_\ell', m_S} & \langle \ell m_\ell; 1 m_S | J m\rangle \,  \langle \ell' m_\ell'; 1 m_S | J' m'\rangle \nonumber \\
& \times \sum_{\bar{\ell}, \bar{m}_\ell} \frac{(4\pi)^{3/2}}{k_\mathsf{cm}^{\bar{\ell} +1} } \, c^{\vec{n}}_{\bar{\ell}, \bar{m}_\ell}(k_\mathsf{cm}^2; L) 
\, \int\!\! d\Omega \; Y_{\ell m_\ell}^*  Y_{\bar{\ell} \bar{m}_\ell}^* Y_{\ell' m_\ell'}    \,,
\end{align}
where the $\text{SU}(2)$ Clebsch-Gordan coefficients encode the $\ell S$ coupling particular to vector-pseudoscalar scattering.
The volume dependence is encoded in the functions $c^{\vec{n}}_{\ell,m_\ell}(k_\mathsf{cm}^2;L)$ which are defined as follows,
\begin{equation}\label{subSlJmj}
c^{\vec{n}}_{\ell,m_\ell}(k_\mathsf{cm}^2;L)=\frac{\sqrt{4\pi}}{\gamma L^3}\bigg(\frac{2\pi}{L}\bigg)^{\ell-2}\,
Z^{\vec{n}}_{\ell,m_\ell}\bigg[1;\bigg(\frac{k_\mathsf{cm}L}{2\pi}\bigg)^2\,\bigg]
\,,\quad
Z^{\vec{n}}_{\ell,m_\ell}[s;x^2]=\sum_{\vec{r}\in \mathcal{P}_{\vec{n}}} \frac{|\vec{r}|^\ell \, Y_{\ell,m_\ell}(\vec{r})}{(|\vec{r}|^2-x^2)^s}\,,
\end{equation}
where the sum is over elements of the set $\mathcal{P}_{\vec{n}} = \big\{\vec{r} \in \mathbb{R}^3 \,|\, \vec{r} = \hat{\gamma}^{-1}(\vec{m}-\alpha\vec{n}) \big\}$, where $\vec{m}$ is an integer triplet, $\vec{n}$ is the normalised vector $\vec{n}=\frac{L}{2\pi}\vec{P}$ as described in Section \ref{secFV}. The scale factor $\alpha=\frac{1}{2}\left[1+\frac{m_1^2-m_2^2}{E^{2}_\mathsf{cm}}\right]$ reflects the asymmetry for unequal masses of scattering particles. $\hat{\gamma}^{-1}$ denotes the Lorentz boost to the centre of momentum frame with $\hat{\gamma}^{-1}\vec{x}\equiv \gamma^{-1}\vec{x}_{\parallel} + \vec{x}_{\perp}$, where $\gamma = E_\mathsf{lab}/E_\mathsf{cm}$ and $\vec{x}_{\parallel}$ and $\vec{x}_{\perp}$ are the components of $\vec{x}$ parallel and perpendicular respectively to the total momentum $\vec{P}$. 

The integral over the product of three spherical harmonics can be expressed in terms of Clebsch-Gordan coefficients,
$$
\int\!\! d\Omega \; Y_{\ell m_\ell}^*  Y_{\bar{\ell} \bar{m}_\ell}^* Y_{\ell' m_\ell'} = \sqrt{ \frac{(2\ell +1)(2\bar{\ell}+1)}{4\pi (2\ell'+1)} } 
\,\langle \ell m_\ell ; \bar{\ell} \bar{m}_\ell | \ell' m_\ell' \rangle \, \langle \ell 0; \bar{\ell} 0 | \ell'0\rangle,
$$ 
and the piece of $\overline{\mathcal{M}}$ independent of the intrinsic spin,
\begin{equation}\label{Ffv}
\sum_{\bar{\ell},\bar{m}_\ell}   \frac{(4\pi)^{3/2}}{k_\mathsf{cm}^{\bar{\ell}+1}}
c^{\vec{n}}_{\bar{\ell}, \bar{m}_\ell}(k_\mathsf{cm}^2;L)
\int \!\! d\Omega \; Y^{*}_{\ell m_\ell} Y^{*}_{\bar{\ell}  \bar{m}_\ell} Y_{\ell' m'_\ell} = F^{\vec{n}}_{\ell m_\ell;\ell'm'_\ell}\, ,
\end{equation}
where $F^{\vec{n}}_{\ell m_\ell;\ell'm'_\ell}$ is the function\footnote{The overall minus sign in Equation (49) of Ref. \cite{Kim:2005gf} is corrected for by the overall minus sign in their definition of $c^{\vec{n}}_{\ell,m_\ell}(k_\mathsf{cm}^2;L)$ -- see Equation (74) of Ref. \cite{Kim:2005gf}.} $F^{FV}_{\ell m_\ell;\ell'm'_\ell}$ in Equation (49) of Ref. \cite{Kim:2005gf} extended to unequal masses by modifying the sum in the generalised zeta functions, $Z^{\vec{n}}_{\ell,m_\ell}$, to be over the set $\mathcal{P}_{\vec{n}}$ defined above -- see Ref. \cite{Leskovec:2012gb}. Furthermore, in Equation (59) of Ref. \cite{Kim:2005gf}, it is shown that
\begin{equation}
F^{\vec{n}}_{\ell m_\ell;\ell'm'_\ell}=i^{\ell'-\ell}\mathcal{M}^{\vec{n}}_{\ell m_\ell;\ell'm'_\ell}
\end{equation}
where $\mathcal{M}^{\vec{n}}_{\ell m_\ell;\ell'm'_\ell}$ is the function defined in Equation (29) of Ref. \cite{Leskovec:2012gb} which is the unequal mass extension to the function $\mathcal{M}^{RG}_{\ell m_\ell;\ell'm'_\ell}$ defined in Equation (89) of Ref. \cite{Rummukainen:1995vs}. In the $S=0$ case the phase-factor $i^{\ell' - \ell}$ cancels completely in the determinant condition and has no effect, while in the present case its effect is felt in e.g. the $\threeSone, \threeDone$ coupled system where different $\ell$ values contribute to the same $J^P$.

The quantisation condition for a given lattice irrep can be obtained by subducing $(J,m)$ components into the irrep $\Lambda$. In the in-flight case, this can be implemented by rotating to a helicity basis and using the helicity-based subductions presented in Table II of~\cite{Thomas:2011rh}. A given $J$ can be subduced into irrep $\Lambda$ more than once, so an \emph{embedding} label, $n$, is required, leaving the space over which the determinant is taken to be $\ell J n$. 

The subduction of $\overline{\mathcal{M}}$ takes the form,
$$
\overline{\mathcal{M}}^{\, \vec{n}, \, \Lambda}_{\, \ell J n, \, \ell' J' n'} \, \delta_{\Lambda,\Lambda'} \delta_{\mu, \mu'} =
\sum_{\substack{m,\, \lambda \\ m',\,  \lambda'}} 
\mathcal{S}^{J\lambda\,*}_{\Lambda \mu n} \, D^{(J)*}_{m \lambda}(R) \;\;
\overline{\mathcal{M}}^{\, \vec{n}}_{\, \ell J m, \, \ell' J' m'} \;\;
\mathcal{S}^{J'\lambda'}_{\Lambda' \mu' n'} \, D^{(J')}_{m' \lambda'}(R)
$$
where $R$ is an active rotation, presented in Table VI of~\cite{Thomas:2011rh}, which takes the $m$ quantisation axis $[001]$ into the direction of $\vec{n}$.

After subduction block-diagonalises into independent irreps, the quantisation condition reads,
\begin{equation}
\det_{\ell J n}\big[ \mathbf{1} + i \,\rho \,\mathbf{t}\,  \big( \mathbf{1} + i \overline{\boldsymbol{\mathcal{M}}}^{\, \vec{n}, \Lambda} \big) \big] = 0 \, ,
\end{equation}
where $\mathbf{1}$ represents $\delta_{\ell, \ell'} \, \delta_{J, J'} \, \delta_{n, n'}$, and where the interpretation of multiple embeddings is that if $J$ is subduced into $\Lambda$ with $N$ embeddings (see Tables~\ref{tab000},~\ref{tab001},~\ref{tab011} and~\ref{tab111}) the $\mathbf{t}$-matrix for that $J$ appears identically as $N$ block diagonal entries in $\mathbf{t}$.

\section{Global fit parameterisations \label{App:Global_Fit_Parameterisations}}
Table~\ref{paramvar} shows the different parameterisations of the $\mathbf{K}$-matrix considered in the parameterisation variation as discussed in detail in Section~\ref{global_analysis}.

\begin{sidewaystable}[h!]
\caption{Polynomial parameterisations of the $\mathbf{K}$-matrix as defined in Eq.~(\ref{Kparams}). Each entry in the table indicates the order of the polynomial, $N(\threelJ | \!\threelprimeJ)$, for the relevant matrix element and ``-'' denotes a zero entry in the $\mathbf{K}$-matrix. The $\chi^2/N_{\text{dof}}$ for each fit, describing the lowest 141 energy levels, is given in the final column. The reference fit, whose parameter values are presented in Table~\ref{ALLparams}, is displayed in bold in the first row of this table.}
\label{paramvar}
{\renewcommand{\arraystretch}{1.3}
\begin{tabular}{ r | ccc :cc: cc : cc | c }
\emph{Phase-space} & $(\threeSone | \threeSone )$ & $(\threeDone | \threeDone )$ & $(\threeSone | \threeDone )$ 
& $(\threePzero | \threePzero )$  & $(\threePone | \threePone )$
& $(\threePtwo | \threePtwo )$  & $(\threePtwo | \threeFtwo )$
& $(\threeDtwo | \threeDtwo )$ & $(\threeDthree | \threeDthree )$ & $\chi^2/N_{\text{dof}}$ \\[0.5ex]
\hline
\multirow{18}{*}{   \rotatebox[origin=c]{90}{Chew-Mandelstam $I(s)$, subtracted at threshold}   }
& {\bf 1} & {\bf 0} & {\bf 0} & {\bf 0} & {\bf 1} & {\bf 1} & {\bf - } & {\bf 0} & {\bf 0} & {\bf 1.42} \\
 & 0 & 0 & 0 & 0 & 0 & 0 & - & 0 & 0 & {\it 2.12} \\
 & 0 & 0 & - & 0 & 0 & 0 & - & 0 & 0 & {\it 2.73} \\
 & 0 & 0 & 0 & 0 & 0 & 0 & 0 & 0 & 0 & {\it 2.12} \\
 & 0 & 0 & 0 & 0 & 0 & 1 & - & 0 & 0 & 1.49 \\
 & 0 & 0 & 0 & 0 & 1 & 1 & - & 0 & 0 & 1.46 \\
 & 1 & 0 & 1 & 0 & 1 & 1 & - & 0 & 0 & 1.42 \\
 & 0 & 0 & 0 & 1 & 1 & 1 & - & 0 & 0 & 1.46 \\
 & 0 & 0 & 0 & 1 & 0 & 0 & - & 0 & 0 & {\it 2.13} \\
 & 0 & 0 & 0 & 0 & 1 & 0 & - & 0 & 0 & {\it 1.82} \\
 & 0 & 0 & 1 & 0 & 0 & 0 & - & 0 & 0 & {\it 2.12} \\
 & 1 & 0 & 1 & 0 & 0 & 1 & - & 0 & 0 & 1.46 \\
 & 2 & 0 & 0 & 0 & 0 & 0 & - & 0 & 0 & {\it 1.96} \\
 & 0 & 0 & 0 & 0 & 1 & 1 & - & 0 & 0 & 1.46 \\
 & 1 & 0 & 1 & 0 & 1 & 1 & - & 0 & 0 & 1.42 \\
 & 1 & 0 & - & 0 & 1 & 1 & - & 0 & 0 & {\it 2.07} \\
 & 1 & 0 & 0 & 0 & 0 & 0 & - & 0 & 0 & {\it 2.05} \\
 & 2 & 0 & 0 & 0 & 1 & 1 & - & 0 & 0 & 1.34 \\
\hdashline
$\mathrm{Re}\, I(s) = 0$
& 1 & 0 & 1 & 0 & 1 & 1 & - & 0 & 0 & 1.44 \\
\end{tabular}
}
\end{sidewaystable}

\bibliographystyle{JHEP}
\bibliography{rhopi_bib}

\end{document}